\documentclass[12pt]{article}

\usepackage[utf8]{inputenc}
\usepackage[T1]{fontenc}
\usepackage[english]{babel}
\usepackage{lmodern}

\usepackage{setspace}
\usepackage[margin=1in]{geometry}
\usepackage{microtype}

\usepackage{tabularx}
\usepackage{longtable}
\usepackage{array}
\usepackage{booktabs}
\usepackage{multirow}
\newcolumntype{L}[1]{>{\raggedright\arraybackslash}p{#1}}
\newcolumntype{R}[1]{>{\raggedleft\arraybackslash}p{#1}}
\newcolumntype{Y}{>{\raggedright\arraybackslash}X}
\usepackage{graphicx}
\graphicspath{{figures/}}
\usepackage{float}
\usepackage[labelfont=bf]{caption}
\usepackage{enumitem}
\setlist{itemsep=0.5ex, topsep=0.5ex, parsep=0.5ex}
\usepackage{xcolor}
\definecolor{sectionblue}{RGB}{0,76,153}
\usepackage{sectsty}
\sectionfont{\color{sectionblue}\large\bfseries\sffamily}
\subsectionfont{\color{sectionblue}\normalsize\bfseries\sffamily}
\subsubsectionfont{\color{sectionblue}\normalsize\sffamily}

\usepackage{amsmath, amssymb, bm}
\DeclareMathOperator*{\argmin}{arg\,min}
\DeclareMathOperator{\logit}{logit}
\DeclareMathOperator{\expit}{expit}
\DeclareMathOperator{\diag}{diag}
\newcommand{\PET}{\mathrm{PET}}
\newcommand{\PEA}{\mathrm{PEA}}
\newcommand{\POP}{\mathrm{POP}}
\newcommand{\TD}{\mathrm{TD}}
\newcommand{\TO}{\mathrm{TO}}
\newcommand{\TGP}{\mathrm{TGP}}
\newcommand{\code}[1]{\texttt{\small #1}}


\usepackage{tikz}
\usetikzlibrary{positioning, arrows.meta, shapes.geometric, fit, calc}
\usepackage{listings}
\definecolor{codebg}{rgb}{0.96,0.96,0.96}
\lstdefinelanguage{toml}{
  comment=[l]{\#},
  morestring=[b]",
  sensitive=true
}

\usepackage[authoryear,round]{natbib}
\usepackage[colorlinks, citecolor=blue, urlcolor=blue, linkcolor=blue, bookmarks=true]{hyperref}

\title{Reconstructing Subnational Labor Indicators in Colombia:\\
An Integrated Machine and Deep Learning Approach}

\author{
  Jaime A. Vera-Jaramillo\thanks{Email: \href{mailto:jaimevera1107@gmail.com}{jaimevera1107@gmail.com},
  \href{mailto:ja.veraj@uniandes.edu.co}{ja.veraj@uniandes.edu.co};
  ORCID: \href{https://orcid.org/0000-0001-7611-6251}{0000-0001-7611-6251}}
}
\date{\today}

\begin{document}
\maketitle

\begin{abstract}
Official labor statistics in Colombia are monthly only at the national level; departments have annual figures, and cities and informality are published as moving windows. This paper presents ICML, an open-source framework that reconstructs monthly series of the labor market, informal employment and employment by sector for the 33 Colombian territories from January 1993 to August 2026. It combines temporal disaggregation, inversion of moving windows, supervised learning with geographic cross-validation and hierarchical reconciliation under exact accounting identities, and publishes each territory for two universes: the whole territory and the part represented by the household survey (GEIH). Informality is obtained from priors of a residual neural network calibrated each month to the national rate. Published values deviate from official figures by a median of 0.04\% to 0.43\%, depending on the quantity, and a leave-one-department-out evaluation gives a mean absolute error of 2.2 percentage points on labor rates. Every record states whether it is observed, anchored, reconstructed or modeled. The results are research estimates, not official statistics.
\end{abstract}

\noindent\textbf{Keywords:} labor statistics; temporal disaggregation; hierarchical reconciliation; small-domain estimation; informality; machine learning; Colombia

\section{Introduction}
\label{sec:introduction}

Labor market monitoring in Colombia rests on the \textit{Gran Encuesta Integrada de Hogares} (GEIH) of the national statistical office (DANE). The survey publishes monthly indicators for the national total, annual indicators for 23 departments, and, for cities, indicators computed over moving windows of several months \citep{DANE_GEIH, DANE_Departamentos}. Informal employment is published for moving quarters \citep{DANE_Informalidad}. No official source provides a monthly labor series for every department, and for nine territories of the Amazon and Orinoco regions and the archipelago of San Andr\'es the survey represents only the capital city, and only since 2014. Before 2001 the survey had a different design and a narrower geographic scope.

Three consequences follow for empirical work. First, a monthly departmental panel does not exist and has to be estimated. Second, the published domains refer to different populations: the national total of the survey excludes part of the national territory, so that departmental figures built on demographic projections cannot add to it. Third, the published figures of different domains are not additive among themselves, so any estimate that respects all of them exactly is impossible.

This paper presents ICML (\textit{Indicadores de Condiciones del Mercado Laboral}), a framework and a software system that reconstructs monthly series for the 33 first-level territories of Colombia from January 1993 to the last month with official publications, August 2026 in the reference vintage described here. Three groups of variables are produced:

\begin{itemize}
  \item Labor market: total population, working-age population, labor force, employed, unemployed, population outside the labor force, and the participation, employment and unemployment rates.
  \item Informality: informal and formal employment and the informality rate.
  \item Employment by sector: employment in 15 groups of economic activity, as an experimental product.
\end{itemize}

Each product is delivered at the national, departmental and city level (the sectoral product has no city level) and for two universes, defined in Section~\ref{sec:framework}: the whole territory and the part of it that the GEIH represents.

The contribution is methodological and operational. On the methodological side, the framework makes explicit four decisions that are usually implicit in reconstructed panels: which population each figure refers to; which published figures are treated as constraints and with what tolerance; who absorbs the inconsistencies between official sources; and which values are observed, anchored to an observation, reconstructed or modeled. On the operational side, the system is designed so that a published figure of a closed year does not change when a new month is added, and so that any result can be rebuilt from the repository and compared with a frozen reference.

The paper does not claim that the model-based components are accurate for the territories without survey coverage: that accuracy cannot be measured, because no observation exists. What is measured, and reported in Section~\ref{sec:results}, is (i) the deviation of the published figures from every official figure that exists, (ii) the error obtained when a territory with official data is treated as if it had none, and (iii) the size of each adjustment the method imposes. Section~\ref{sec:limitations} lists what remains unverified.

The remainder of the paper is organized as follows. Section~\ref{sec:background} describes the statistical sources and the related literature. Section~\ref{sec:framework} defines the universes, the notation and the structure of the pipeline. Sections~\ref{sec:labor}, \ref{sec:informality} and \ref{sec:sectoral} describe the estimation of the labor market, informality and sectoral employment. Section~\ref{sec:system} describes the production system: vintages, gates and reproducibility. Section~\ref{sec:results} reports the validation evidence and the resulting series. Section~\ref{sec:limitations} discusses limitations and Section~\ref{sec:conclusions} concludes.

\section{Statistical Sources and Related Work}
\label{sec:background}

\subsection{What the official sources publish}
\label{sec:sources}

Table~\ref{tab:sources} lists the sources used and the temporal support of each one. The distinction between a calendar month, a moving window and a calendar year is maintained throughout: a figure published for a window constrains that window and is never treated as an observation of a month.

\begin{table}[!htbp]
\centering
\caption{Sources and their published support.}
\label{tab:sources}
\footnotesize
\begin{tabularx}{\textwidth}{L{3.6cm} L{3.2cm} Y}
\toprule
\textbf{Source} & \textbf{Support} & \textbf{Role in the framework} \\
\midrule
GEIH monthly labor-market annex \citep{DANE_GEIH} & National, monthly, from 2001-01 & National reference totals of the six labor quantities; sets the cutoff month together with the informality annex \\
GEIH monthly annex, city sheet & 32 cities, 12-month moving windows, from the window ending 2021-12 & Linear constraints on latent monthly urban values (57 windows at the August 2026 cutoff) \\
GEIH annual departmental annex \citep{DANE_Departamentos} & 23 departments, annual, 2007--2025 & Annual anchors of the anchored departments; defines the years that an ordinary update cannot modify \\
Informality annexes \citep{DANE_Informalidad} & 13 areas, 23 cities and national total; 3-month moving windows & National and urban informality targets, by definitional regime (Section~\ref{sec:informality}) \\
ENH household survey & National, annual, 1993--2001 & Annual levels of the pre-GEIH history \\
Monthly price indicators (consumer and producer price indices, UVR) & National, monthly, 1993--2003 & Related series for the temporal disaggregation of the pre-GEIH history \\
Population projections \citep{DANE_Proyecciones} & Departments and municipalities, annual & Whole-territory population of each department; capital and rest of territory \\
CNPV 2018 census \citep{DANE_CNPV2018} & Municipal heads of nine capitals & Anchor of the bridge between universes \\
Sectoral annexes and departmental accounts \citep{DANE_CuentasDepartamentales} & National monthly (13 groups, from 2015); departmental annual shares (23 departments, 2015--2025); departmental GDP & Sectoral composition and predictors (Section~\ref{sec:sectoral}) \\
\bottomrule
\end{tabularx}
\end{table}

Three features of these sources shape the method.

The national total of the GEIH excludes, in the Amazon and Orinoco departments, the municipal heads other than the capital, the populated centers and the dispersed rural area, as well as Providencia and rural San Andr\'es; the capitals of those departments and San Andr\'es enter the national total from 2014. With official figures only, the national population of 2022--2025 equals the sum of the 23 departments of the annual annex, Bogot\'a and the nine capitals within 0.013--0.018\% ($-0.026\%$ in 2021), and the working-age population within 0.02\%. Between 2007 and 2013 the gap between the sum of departmental demographic projections and the national GEIH population is 2.81--2.88\%, which matches the population share of the nine territories (2.77--2.89\%), and it falls to 2.12\% in 2014. The gap is therefore survey coverage, not estimation error.

The published domains do not add up. Table~\ref{tab:nonadditivity} shows the difference between the national total and the sum of its official parts.

\begin{table}[H]
\centering
\caption{Sum of official domains (23 departments, Bogot\'a and nine capitals) minus the official national total, as a percentage of the national total.}
\label{tab:nonadditivity}
\small
\begin{tabular}{lrrrr}
\toprule
\textbf{Year} & \textbf{Employed} & \textbf{Unemployed} & \textbf{Outside labor force} & \textbf{Population} \\
\midrule
2021 & $+0.41$ & $-1.90$ & $-0.18$ & $-0.026$ \\
2022 & $-0.10$ & $-0.30$ & $+0.27$ & $+0.018$ \\
2023 & $-0.07$ & $-0.66$ & $+0.30$ & $+0.013$ \\
2024 & $+0.06$ & $-0.70$ & $+0.09$ & $+0.014$ \\
2025 & $-0.04$ & $-0.28$ & $+0.16$ & $+0.014$ \\
\bottomrule
\end{tabular}
\end{table}

Population is practically additive and unemployment is not; 2021 is the year in which the sampling frame of the survey changed. Any estimator that reproduces the national total has to assign this residual to someone (Section~\ref{sec:representativeness}).

Three discontinuities exist in the measurement itself: the change from the ENH to the continuous survey in 2001, the redesign of the GEIH with a new sampling frame in 2021, and the change in the operational definition of informality in 2021. The framework never joins two regimes by assuming that their definitions are equal; each seam is treated with a documented rule (Sections~\ref{sec:reconstruction} and \ref{sec:informality}).

\subsection{Related work}
\label{sec:literature}

\subsubsection{Temporal disaggregation and benchmarking}

The problem of deriving a high-frequency series consistent with low-frequency totals goes back to \citet{Denton1971}, who adjusts a preliminary series by quadratic minimization, and to \citet{ChowLin1971}, who formulate it as generalized least squares with related series and autoregressive errors. \citet{Fernandez1981} and \citet{Litterman1983} extend the error model to non-stationary residuals, \citet{DagumCholette2006} give a unified treatment of benchmarking and reconciliation, and \citet{Mosley2022} address the case with many candidate indicators. The implementation used here is the \code{tempdisagg} library \citep{VeraJaramillo2025tempdisagg}.

\subsubsection{Reconciliation of accounts and hierarchies}

Adjusting a set of estimates so that they satisfy accounting identities, with adjustments proportional to the assumed reliability of each estimate, was formalized by \citet{StoneChampernowneMeade1942}. Iterative proportional fitting \citep{DemingStephan1940} and calibration estimators \citep{DevilleSarndal1992} solve the related problem of adjusting a table or a set of weights to known margins with minimum distance. In the forecasting literature, the same structure appears as the reconciliation of hierarchical time series \citep{Hyndman2011, Wickramasuriya2019}. The closures described in Sections~\ref{sec:labor}--\ref{sec:sectoral} are instances of these ideas: quadratic programs with exact linear constraints for levels, and minimum-divergence adjustments for rates and shares.

\subsubsection{Estimation for domains without direct observations}

Small area estimation combines direct survey estimates with models that borrow strength from auxiliary variables \citep{FayHerriot1979, RaoMolina2015}. Recent applications produce district-level maps of labor outcomes by linking a household survey with census covariates \citep{Srivastava2025}. A related line uses machine learning to downscale labor statistics to areas where no survey estimate exists \citep{vanDijk2022}. The present setting differs from the standard small-area problem in that the microdata are not used: the inputs are published aggregates with heterogeneous temporal support, and the territories of interest include some for which no direct estimate has ever been published. The framework therefore treats the supervised components as one layer among several, evaluates them only where official figures exist, and reports that evaluation after the reconciliation step, which is the form in which the estimates are actually published.

\subsubsection{Auxiliary signals and machine learning for labor indicators}

A separate literature uses high-frequency auxiliary signals to anticipate labor indicators before the official release. Internet search volumes improve short-term forecasts of unemployment \citep{AskitasZimmermann2009, ChoiVarian2012, DAmuriMarcucci2017}, and behavioral traces from social media correlate with regional unemployment \citep{Llorente2015}. For Colombia, \citet{CardonaRojas2017} forecast the national unemployment rate with search data, \citet{OrozcoCastaneda2024} combine macroeconomic variables, labor metrics and search data in a machine-learning indicator of the national labor market, \citet{VidalSierraCeron2025} build a regional indicator for four cities with variable selection by regularization, and \citet{PerezRosero2025} predict the national unemployment rate with Gaussian processes and an interpretable low-dimensional embedding. These studies address forecasting or nowcasting of a published series. The problem treated here is different: the target is a complete panel of levels and rates for territories and periods in which no series is published, subject to accounting identities and to official totals.

\subsubsection{Colombian labor market}

The series produced here cover episodes that have been studied with national or urban data, among them the recession of 1999 the global financial crisis of 2008--2009 \citep{Grosh2014} and the contraction and informality-led recovery of 2020--2021 \citep{AlvarezPizzinelli2021}. Other studies with survey microdata evaluate specific labor policies \citep{MedinaNunezTamayo2013} or measure occupational exposure to new technologies in a group of countries that includes Colombia \citep{Pizzinelli2023}. The purpose of this paper is not to reinterpret those episodes but to document a dataset that allows them to be examined at the departmental level with explicit provenance.

\section{Framework}
\label{sec:framework}

\subsection{Units, quantities and identities}

The unit of analysis is the territory-month. Let $d \in \mathcal{D}$ index the 33 first-level territories (32 departments and Bogot\'a D.C.) and $t = 1,\dots,T$ the months from January 1993 to the cutoff ($T = 404$ at August 2026). For each territory-month the labor product contains six levels, in persons: population $\POP_{d,t}$, working-age population $\PET_{d,t}$, labor force $\PEA_{d,t}$, employed $E_{d,t}$, unemployed $U_{d,t}$ and population outside the labor force $I_{d,t}$. They satisfy
\begin{align}
\PEA_{d,t} &= E_{d,t} + U_{d,t}, \label{eq:id-pea}\\
\PET_{d,t} &= \PEA_{d,t} + I_{d,t}, \label{eq:id-pet}\\
\PET_{d,t} &\le \POP_{d,t}, \label{eq:id-pop}
\end{align}
and the rates are derived from the levels:
\begin{equation}
\TGP_{d,t} = \frac{\PEA_{d,t}}{\PET_{d,t}}, \qquad
\TO_{d,t} = \frac{E_{d,t}}{\PET_{d,t}}, \qquad
\TD_{d,t} = \frac{U_{d,t}}{\PEA_{d,t}},
\label{eq:rates}
\end{equation}
so that $\TO = \TGP\,(1-\TD)$ holds identically. Informal and formal employment, $F^{\mathrm{inf}}_{d,t}$ and $F^{\mathrm{for}}_{d,t}$, satisfy
\begin{equation}
E_{d,t} = F^{\mathrm{inf}}_{d,t} + F^{\mathrm{for}}_{d,t},
\label{eq:id-inf}
\end{equation}
and employment by sector $E_{d,s,t}$, $s \in \mathcal{S}$ with $|\mathcal{S}| = 15$, satisfies $\sum_{s} E_{d,s,t} = E_{d,t}$. Identities \eqref{eq:id-pea}, \eqref{eq:id-pet} and \eqref{eq:id-inf} are enforced within $10^{-5}$ persons in the labor product and within $10^{-3}$ persons across products (Section~\ref{sec:gates}).

Cities are a separate set of 33 units: 32 cities or metropolitan areas of the GEIH and one synthetic urban signal for Cundinamarca. City series are \emph{signals}: they are not additive, they do not partition the country and they are never summed.

\subsection{Two universes}
\label{sec:universes}

The departmental demographic projections cover the whole territory. The national total of the GEIH covers less (Section~\ref{sec:sources}). Rather than forcing the departments to add to a national total that excludes part of them, the framework publishes every territory for two universes:

\begin{itemize}
  \item Census universe: the whole territory of each department, consistent with the demographic projections based on the 2018 census. The ICML national total is defined as the unweighted sum of the 33 territories:
  \begin{equation}
  X^{\mathrm{ICML}}_{t} = \sum_{d \in \mathcal{D}} X_{d,t}, \qquad X \in \{E, U, I, \PEA, \PET, \POP\}.
  \label{eq:icml-national}
  \end{equation}
  \item GEIH universe: the part of each territory that the survey represents, $X^{\mathrm{rep}}_{d,t} = \phi^{X}_{d,t}\,X_{d,t}$, where $\phi^{X}_{d,t} \in [0,1]$ is a representativeness factor defined per territory, month and quantity (Section~\ref{sec:representativeness}).
\end{itemize}

\begin{figure}[H]
\centering
\includegraphics[width=0.95\textwidth]{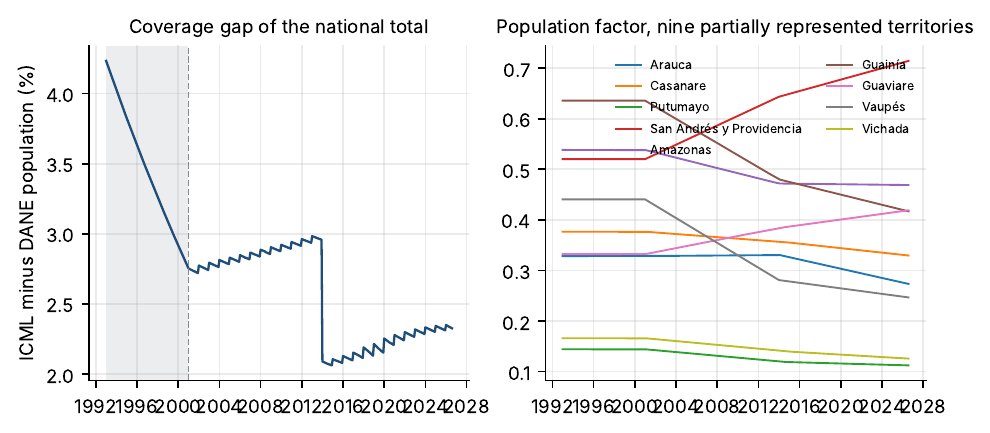}
\caption{Left: excess of the ICML national population over the DANE reference. From 2001 it lies between 2.06\% and 2.98\%; the step in 2014 corresponds to the incorporation of the nine capitals into the national total of the survey. Right: population factor of the nine partially represented territories.}
\label{fig:coverage}
\end{figure}

Two national series are therefore kept side by side and never added: the ICML national total \eqref{eq:icml-national} and the DANE reference, which from January 2001 is the published national GEIH series and before that date is a reconstructed reference flagged as non-official. The ICML national population exceeds the DANE reference by 2.06--2.98\% between 2001 and 2026 (Figure~\ref{fig:coverage}).

\subsection{Classes of evidence}
\label{sec:evidence}

Every record carries its provenance. Four classes of values are distinguished, and the distinction is part of the published tables:

\begin{enumerate}
  \item Observed: a figure published by DANE for exactly that unit and period (national months from 2001; annual departmental figures; window means).
  \item Anchored: a monthly value estimated subject to an exact or bounded constraint from an observed figure (months of an anchored department; latent months of a city constrained by its windows).
  \item Reconstructed: a value for a period without official observation of that unit, derived from another regime or another unit by a deterministic rule (the 1993--2000 history; donor-based urban histories).
  \item Modeled: a value produced by a fitted statistical model without a constraint of its own (rest-of-territory ratios; informality priors; departmental sector shares).
\end{enumerate}

\subsection{Structure of the pipeline}

Figure~\ref{fig:pipeline} shows the chain. A run executes nine processes in a fixed order; a blocking failure in any of them stops the run without publishing:

\begin{enumerate}
  \item Reception and planning: the annexes are detected, the cutoff is set at the last month for which both monthly annexes exist, and each workbook is verified to declare internally the month its name states.
  \item Source preparation: nine extractors convert the workbooks into audited tables with recorded hashes.
  \item Labor chain: seven stages (national, urban, departmental, integration, modeling, reconciliation and quality; Section~\ref{sec:labor}).
  \item Labor reconstruction 1993--2000 (Section~\ref{sec:reconstruction}).
  \item Regeneration of the informality sequence from the annexes and the current labor product.
  \item Informality closure (Section~\ref{sec:informality}).
  \item Sectoral estimation (Section~\ref{sec:sectoral}).
  \item Cross audit of the eight products and comparison with the reference vintage.
  \item Consolidation and delivery by universe (Section~\ref{sec:system}).
\end{enumerate}

\begin{figure}[H]
\centering
\begin{tikzpicture}[
  node distance=5mm and 7mm,
  box/.style={draw, rounded corners=2pt, align=center, font=\scriptsize\sffamily, minimum height=8mm, text width=27mm, fill=blue!5},
  src/.style={box, fill=gray!12},
  prod/.style={box, fill=green!10},
  arr/.style={-{Stealth[length=2mm]}, thick, gray!70!black}
]
\node[src] (ann) {DANE annexes and frozen parameters};
\node[box, below=of ann] (prep) {Source preparation (9 extractors)};
\node[box, below left=7mm and -6mm of prep] (nat) {National stage};
\node[box, below right=7mm and -6mm of prep] (urb) {Urban stage};
\node[box, below=of nat] (dep) {Departmental stage};
\node[box, below=of urb] (int) {Integration (latent months)};
\node[box, below=of dep] (mod) {Modeling};
\node[box, below=of int] (rec) {Reconciliation and factors};
\node[box, below=10mm of $(mod.south)!0.5!(rec.south)$] (qual) {Quality gate and 1993--2000 reconstruction};
\node[prod, below=of qual] (lab) {Labor: nationals, departments, cities};
\node[box, left=of lab] (infq) {Informality sequence and closure};
\node[box, right=of lab] (sec) {Sectoral predictor and raking};
\node[prod, below=of infq] (infp) {Informality: 3 levels};
\node[prod, below=of sec] (secp) {Sectoral: national, departmental};
\node[box, below=14mm of lab] (aud) {Cross audit and reference comparison};
\node[prod, below=of aud] (del) {Delivery by universe (census, GEIH)};
\draw[arr] (ann) -- (prep);
\draw[arr] (prep) -- (nat);
\draw[arr] (prep) -- (urb);
\draw[arr] (nat) -- (urb);
\draw[arr] (nat) -- (dep);
\draw[arr] (urb) -- (int);
\draw[arr] (dep) -- (int);
\draw[arr] (dep) -- (mod);
\draw[arr] (int) -- (mod);
\draw[arr] (mod) -- (rec);
\draw[arr] (int) -- (rec);
\draw[arr] (rec) -- (qual);
\draw[arr] (qual) -- (lab);
\draw[arr] (lab) -- (infq);
\draw[arr] (lab) -- (sec);
\draw[arr] (infq) -- (infp);
\draw[arr] (sec) -- (secp);
\draw[arr] (lab) -- (aud);
\draw[arr] (infp) -- (aud);
\draw[arr] (secp) -- (aud);
\draw[arr] (aud) -- (del);
\end{tikzpicture}
\caption{Structure of the pipeline. Gray: inputs; blue: processes; green: the eight internal products (three labor, three informality, two sectoral).}
\label{fig:pipeline}
\end{figure}

The parameters of every stage are declared in versioned configuration files; the values quoted in the following sections are those of the reference vintage and are collected in Appendix~\ref{app:parameters}.

\section{Labor Market Estimation}
\label{sec:labor}

The labor chain has seven stages. Each stage is a pure computation that reads the artifacts of previous stages and the prepared sources, and writes its outputs once, with the SHA-256 hash of every input and output. This section describes what each stage estimates and under which constraints.

\subsection{National stage}
\label{sec:national}

From January 2001 the national series are the published GEIH months and are not modified. The stage builds the months from January 1993 to December 2000 and attaches them to the observed series.

\subsubsection{Annual splice}

The annual levels of the ENH (1993--2001) and of the GEIH are joined, for each quantity separately, with a ratio splice over two overlapping periods: the earlier series is multiplied by the mean ratio between the two series in the overlap.

\subsubsection{Temporal disaggregation}

Each annual series $\bm{y}_a \in \mathbb{R}^{n}$ is distributed to months with the regression method of \citet{ChowLin1971}. Let $\bm{y} \in \mathbb{R}^{12n}$ be the unknown monthly series, $\bm{C}$ the $n \times 12n$ matrix that averages months into years, and $\bm{X}$ a matrix with an intercept and a related monthly indicator:
\begin{equation}
\bm{y} = \bm{X}\bm{\beta} + \bm{u}, \qquad u_t = \rho\,u_{t-1} + \varepsilon_t, \qquad \bm{y}_a = \bm{C}\bm{y}.
\label{eq:chowlin-model}
\end{equation}
With $\bm{V}(\rho)$ the covariance matrix of $\bm{u}$, the estimator is
\begin{align}
\hat{\bm{\beta}} &= \left[\bm{X}^{\top}\bm{C}^{\top}(\bm{C}\bm{V}\bm{C}^{\top})^{-1}\bm{C}\bm{X}\right]^{-1}\bm{X}^{\top}\bm{C}^{\top}(\bm{C}\bm{V}\bm{C}^{\top})^{-1}\bm{y}_a, \label{eq:chowlin-beta}\\
\hat{\bm{y}} &= \bm{X}\hat{\bm{\beta}} + \bm{V}\bm{C}^{\top}(\bm{C}\bm{V}\bm{C}^{\top})^{-1}\left(\bm{y}_a - \bm{C}\bm{X}\hat{\bm{\beta}}\right), \label{eq:chowlin-y}
\end{align}
and $\rho$ is chosen by maximizing the Gaussian log-likelihood of the annual residuals. A solution with $\rho$ on the boundary of its domain is rejected. The indicators are fixed by configuration: the real value unit (UVR) for employed, the producer price index for the population outside the labor force and the consumer price index \citep{DANE_IPC} for the remaining quantities. These indicators carry little labor-market information; their role is to provide a smooth, strictly positive monthly profile, and the dynamics of the reconstructed history are set later (Section~\ref{sec:reconstruction}).

\subsubsection{Accounting reconciliation}

Because each quantity is disaggregated independently, the annual levels of the history do not satisfy identities \eqref{eq:id-pea}--\eqref{eq:id-pet}. For each historical year (through 2000), the vector $\bm{a} \in \mathbb{R}^{6}$ of the six annual levels is projected onto the subspace of identity-consistent vectors by weighted least squares in the spirit of \citet{StoneChampernowneMeade1942}. Writing $\bm{a} = \bm{M}\bm{b}$ for a consistent vector generated by four free components $\bm{b} = (E, U, I, N)^{\top}$, with $N$ the population outside working age,
\begin{equation}
\hat{\bm{b}} = \argmin_{\bm{b}} \sum_{k=1}^{6} \left(\frac{a_k - (\bm{M}\bm{b})_k}{\sigma_k}\right)^{2}, \qquad \sigma_k = \max\left(0.05\,|a_k|,\ 10^{4}\right),
\label{eq:national-wls}
\end{equation}
so that every anchor is assumed to have a relative uncertainty of 5\% with a floor of ten thousand persons. Observed GEIH years are not adjusted; only their derived quantities are recomputed from the base components to remove publication rounding (at most one person). The monthly series are then rescaled proportionally to the reconciled annual means. The stage stops if a year moves by more than 10\% or a month by more than 15\%.

\subsubsection{Seasonal transfer}

The reconstructed months inherit the weak seasonality of their indicators. With a robust STL decomposition \citep{Cleveland1990}, the stage estimates the log-seasonal profile of each component share ($E/\PET$, $U/\PET$, $I/\PET$) in the history and in an observed reference window (2004--2019), applies a fraction $\kappa = 0.5$ of the difference between both profiles to the history, and restores the monthly working-age population and the annual mean of every component with iterative proportional fitting. Months from 2001 onward are not touched.

\subsection{Urban stage}
\label{sec:urban}

The urban stage produces monthly signals for the cities from January 1993 to December 2020. For 23 cities the GEIH provides annual levels from 2007 to 2020. For the capitals of the nine partially represented territories (Arauca, Yopal, Mocoa, San Andr\'es, Leticia, In\'irida, San Jos\'e del Guaviare, Mit\'u and Puerto Carre\~no) no urban history exists before the 12-month windows that start in 2021.

\subsubsection{Donor selection}

A recipient city $r$ borrows the \emph{trajectory} of a donor city $g$ with history, rescaled to its own level:
\begin{equation}
y_{r}(\tau) = \alpha_{r,g}\, y_{g}(\tau), \qquad \tau = 2007,\dots,2020,
\label{eq:donor}
\end{equation}
where $\alpha_{r,g}$ is calibrated on the years 2021--2024 of the annual aggregates of the published windows, and 2025 is held out. Donors are chosen by a Pareto analysis over five loss criteria (growth of employment, growth of population, rates, scale and stability of the ratio), a weighted-regret rule to break ties among non-dominated candidates, and two guards relative to the incumbent donor: the error in the held-out year and the continuity at the seam of 2021. The rule was validated retrospectively on the 23 cities with known history, by treating each one as a recipient. The selected donors are frozen: later observations never change them.

\subsubsection{Bounded closure}

The transferred history of a recipient may not connect with its first observed window. The seam residual of each component is shrunk to a target mean in logarithms, capped, and distributed as a log-linear ramp over 2015--2020. Values from 2021 onward and years before the ramp are not modified.

\subsubsection{Monthly profile and accounting projection}

For each city and year from 2007 to 2020, the monthly profile is that of the homologous national series, rescaled to the annual anchor. Because six signals are generated from four free non-negative components, they are projected onto the accounting subspace by least squares on relative errors:
\begin{equation}
\hat{\bm{b}} = \argmin_{\bm{b} \ge 0} \left\| \diag(\bm{w})^{-1}\left(\bm{s} - \bm{M}\bm{b}\right) \right\|_2^{2},
\label{eq:urban-projection}
\end{equation}
where $\bm{s}$ are the six signals and $\bm{w}$ a reference vector fixed within each block, which makes the projection preserve the mean of anchors that are already coherent. Before 2007 the national series is linked to the 2007 level with a ratio splice over 12 overlapping months.

\subsubsection{Synthetic signal for Cundinamarca}

Cundinamarca has no urban domain in the survey. Its signal uses Bogot\'a only as a temporal pattern, scaled to Cundinamarca's own departmental anchors over 2021--2024 with a ratio splice. It is a predictor for later stages and is labeled as synthetic in every product.

\subsection{Departmental stage}
\label{sec:departmental}

\subsubsection{Annual anchors}

The annual annex provides, for 23 departments and for 2007--2025, the six levels and the three rates. Parsing is fail-closed: an unknown title, a repeated year, a missing concept or a non-numeric cell is an error and nothing is imputed. Labor force and working-age population are recomputed from their components; if the change exceeds the publication rounding (2.01 persons), the source is declared inconsistent and the stage stops.

\subsubsection{Monthly reconstruction}

For each anchored department and each free component (employed, unemployed, outside the labor force, population), the annual means are distributed to months with \eqref{eq:chowlin-model}--\eqref{eq:chowlin-y}, using the homologous national monthly series as the related indicator. The annual means are reproduced within $10^{-4}$ persons. Before 2007, the departmental series is linked to the national series with a ratio splice over 12 months; labor force and working-age population follow from the identities. For one series (the population outside the labor force of Cundinamarca) the likelihood has two local maxima in $\rho$; the global maximum is kept and the monthly difference with respect to the competing solution is recorded as a sensitivity audit.

\subsubsection{Demographic backcast of population}

Before 2007 the reconstruction would inherit the national population dynamics, which can differ from the demography of each territory. The pre-2007 population is therefore replaced by the department's demographic projection, rescaled to its official GEIH level in 2007, interpolated to months with a monotone piecewise cubic curve \citep{FritschCarlson1980} and rebalanced so that every annual target is reproduced exactly.

\subsection{Integration stage: latent months from moving windows}
\label{sec:integration}

From the window ending in December 2021, DANE publishes for 32 cities the mean of each labor quantity over 12-month moving windows. With $K$ windows there are $K + 11$ months involved; at the August 2026 cutoff, 57 windows and 68 months (January 2021 to August 2026). Let $\bar{w}_k$ be the published mean of window $k$ and $y_m$ the unknown value of month $m$. Consecutive windows differ by one month entering and one leaving, so
\begin{equation}
y_{m} = y_{m-12} + 12\left(\bar{w}_{k} - \bar{w}_{k-1}\right),
\label{eq:window-recursion}
\end{equation}
and all months are determined by twelve free base values $\bm{z} \in \mathbb{R}^{12}$:
\begin{equation}
\bm{y} = \bm{H}\bm{z} + \bm{b}, \qquad \frac{1}{12}\,\bm{1}^{\top}\bm{z} = \bar{w}_1,
\label{eq:window-affine}
\end{equation}
where $\bm{H}$ assigns each month to its calendar position and $\bm{b}$ accumulates the differences \eqref{eq:window-recursion}. Any $\bm{z}$ that satisfies the constraint reproduces every window exactly. The free values are chosen by a strictly convex quadratic program:
\begin{equation}
\min_{\bm{z}} \;
\left\| \bm{y}(\bm{z}) - \bm{p} \right\|_2^{2}
+ \lambda_{s} \left\| \bm{D}_1\left(\bm{y}(\bm{z}) - \bm{p}\right) \right\|_2^{2}
+ \lambda_{b} \left( z_{1} - y_{0}\,\frac{n_{1}}{n_{0}} \right)^{2}
\quad \text{s.t. } \eqref{eq:window-affine},\; \bm{z} \ge \underline{\bm{z}},
\label{eq:window-qp}
\end{equation}
with all quantities expressed in units of the first window mean. Here $\bm{p}$ is a prior equal to the national homologous series scaled by the ratio between the city and national window means; $\bm{D}_1$ is the first-difference operator, so the second term keeps the month-to-month changes close to those of the prior ($\lambda_s = 0.65$); and the third term softly links January 2021 ($z_1$) to the approved value of December 2020 ($y_0$) carried forward with the national change between both months ($n_1/n_0$), with $\lambda_b = 0.12$. The lower bound $\underline{\bm{z}}$ keeps every month above 10\% of the smallest window, which prevents algebraic compensation from producing near-zero months; when the unconstrained solution violates it, the bounded problem is solved by sequential quadratic programming. These months are latent estimates and are never labeled as monthly observations.

\subsubsection{Control of implausible jumps}

Latent months reproduce the windows exactly but can oscillate for small cities, where the windows carry little information about individual months. For unemployment in the latent segment, two scenarios are built. In the \emph{strict} scenario the windows remain exact and the month-to-month ratio is bounded by the smallest bidirectional cap compatible with them, not smaller than 2. In the \emph{quality-filtered} scenario, which is the one published, cities that need no regularization are left unchanged, regularized cities take the strict solution, and three small capitals (Leticia, In\'irida and Mit\'u), for which no finite cap is attainable with exact windows, are allowed a bounded window tolerance of at most 2.1\%. That tolerance is a modeling assumption and not a sampling margin.

\subsubsection{Propagation}

The seasonal transfer of the national history (Section~\ref{sec:national}) is propagated to the urban and departmental series for 1993--2000 by reapplying their link factors to the adjusted national series. Nothing is re-estimated, months from 2001 remain identical and annual means do not change.

\subsection{Supervised models of departmental rates}
\label{sec:modeling}

The modeling stage fits models that predict departmental rates from urban, national and demographic information. Their role in the published product is limited and is stated at the end of this subsection.

\subsubsection{Targets}

Four rates are modeled independently: $\TO$, $\TD$, $\TGP$ and the ratio of working-age population to the demographic projection. Predictions are projected onto the accounting plane $\TO = \TGP\,(1 - \TD)$. Labels are the monthly paths of the anchored departments, which are constrained to annual anchors and are not monthly observations; evaluation is therefore done on annual means. The training period is 2007--2025. Cundinamarca is excluded from the labeled set, because its synthetic urban signal is calibrated with its own anchors.

\subsubsection{Predictors}

Predictors come only from the urban signals, the national series and the demographic projections; departmental labels are kept in a separate table and never enter predictor generation. The base set has thirteen variables: the urban unemployment, employment and participation rates; the national unemployment rate; the differences between urban and national unemployment and employment rates; the logarithm of the projected departmental population; the ratio of urban population to the departmental projection; the year-on-year growth of urban employment and population (clipped to $[-90\%, 250\%]$); two harmonic terms of the month; and the year. Lags of 1 and 12 months are added.

\subsubsection{Territorial clusters}

Each territory is summarized by a profile of medians of urban rates and structure, plus growth, built with information up to a cutoff year. Cluster memberships from $k$-means or from the hierarchical method of \citet{Ward1963} enter as one-hot features. The scaler, the partition and the number of clusters ($k \le 5$) are refitted in every fold and in the final fit with training territories only.

\subsubsection{Validation design}

The 22 labeled non-synthetic territories are split into five groups with a seeded permutation. Each fold holds out one group for a future window (2016--2019, 2019--2022 or 2022--2025), trains on the other groups up to the cutoff, and drops the last training year as an embargo, since annualized 12-month windows would otherwise overlap the test period. This gives 15 geographic-temporal folds. Hyperparameters are fixed before the comparison and the test fold is never used for tuning or early stopping.

\subsubsection{Candidates}

Five methods are compared: the urban analog (the urban rate itself), ridge regression \citep{HoerlKennard1970} with penalty 25, gradient-boosted trees \citep{ChenGuestrin2016} without clusters, and gradient-boosted trees with $k$-means or Ward memberships. The boosted models use 200 trees of depth 3, learning rate 0.045, minimum child weight 12, row and column subsampling 0.9, $L_2$ penalty 8 and $L_1$ penalty 0.05, with the histogram algorithm and a single thread. Four hybrid portfolios, declared before evaluation, take each rate from a given source; the main challenger takes $\TO$ and $\TGP$ from the urban signal and $\TD$ and the working-age ratio from the boosted model.

\subsubsection{Role in the published product}

The stage produces a 33-territory panel in which the anchored departments keep their anchored reconstruction for every year with a published annex. For the months after the last anchored year (2026 in the reference vintage) the anchored departments take the model path, linked to the anchored segment with a 12-month taper. Territories without an annual anchor are \emph{not} taken from this panel in the published product: Bogot\'a and the nine partially represented territories are built from their own survey domains (Section~\ref{sec:representativeness}). The evaluation that motivated this design is reported in Section~\ref{sec:transfer}.

\subsection{Reconciliation and representativeness factors}
\label{sec:representativeness}

The published departmental product is built from three kinds of represented parts and one estimated complement.

\subsubsection{Represented parts}

(i) For the 23 anchored departments, the anchored monthly reconstruction, with the population of the departmental stage. (ii) For Bogot\'a, its urban domain in the survey, which represents the whole territory. (iii) For each of the nine partially represented territories, the urban signal of its capital. Let $R_{u,t}^{X}$ denote the represented part of unit $u$ for component $X \in \{E, U, I, \POP\}$.

\subsubsection{Allocation of the national residual}

From January 2001, the represented parts are reconciled to the published national total $X^{\mathrm{DANE}}_t$. Because the official domains are not additive (Table~\ref{tab:nonadditivity}), the residual
\begin{equation}
r^{X}_{t} = X^{\mathrm{DANE}}_{t} - \sum_{u \in \mathcal{U}_t} R^{X}_{u,t}
\label{eq:residual}
\end{equation}
is distributed in proportion to the size of each part:
\begin{equation}
\tilde{R}^{X}_{u,t} = R^{X}_{u,t} + \frac{\bar{R}^{X}_{u,y(t)}}{\sum_{v \in \mathcal{U}_t}\bar{R}^{X}_{v,y(t)}}\; r^{X}_{t},
\label{eq:proportional}
\end{equation}
where $\bar{R}^{X}_{u,y}$ is the mean of the part in year $y$ and $\mathcal{U}_t$ is the set of parts inside the national total in month $t$ (a capital of the nine territories enters only from January 2014). The persons absorbed by each territory, $\tilde{R}^{X}_{u,t} - R^{X}_{u,t}$, are published in a column of their own, so that the official figure each estimate started from can be recovered. The alternative of keeping every annual anchor exact and assigning the whole residual to the units without an anchor was implemented and rejected on the evidence of Section~\ref{sec:residual-evidence}.

\subsubsection{Whole-territory levels}

For anchored departments and Bogot\'a the whole territory equals the represented part. For the nine partially represented territories,
\begin{equation}
X_{d,t} = X^{\mathrm{cap}}_{d,t} + X^{\mathrm{rest}}_{d,t},
\label{eq:capital-rest}
\end{equation}
where the population of the rest is the departmental projection minus the population of the capital, and its labor levels follow from three ratios:
\begin{align}
\POP^{\mathrm{rest}}_{d,t} &= \POP^{\mathrm{proj}}_{d,t} - \POP^{\mathrm{cap}}_{d,t}, &
\PET^{\mathrm{rest}}_{d,t} &= q^{\PET}_{y(t)}\,\POP^{\mathrm{rest}}_{d,t}, \nonumber\\
\PEA^{\mathrm{rest}}_{d,t} &= q^{\TGP}_{y(t)}\,\PET^{\mathrm{rest}}_{d,t}, &
U^{\mathrm{rest}}_{d,t} &= q^{\TD}_{y(t)}\,\PEA^{\mathrm{rest}}_{d,t},
\label{eq:rest}
\end{align}
with $E^{\mathrm{rest}} = \PEA^{\mathrm{rest}} - U^{\mathrm{rest}}$ and $I^{\mathrm{rest}} = \PET^{\mathrm{rest}} - \PEA^{\mathrm{rest}}$. The ratios $q_y$ are the unweighted means, across anchored departments (excluding Cundinamarca), of the ratios observed in their own rest of territory, computed as department minus capital on annual means; years without donor data carry the last available year forward. The same rule is used for all nine territories. Section~\ref{sec:rest-evidence} reports its validation.

\subsubsection{Factors}

The representativeness factor of quantity $X$ is the ratio between the represented part and the whole territory,
\begin{equation}
\phi^{X}_{d,t} = \frac{X^{\mathrm{rep}}_{d,t}}{X_{d,t}},
\label{eq:factor}
\end{equation}
which equals 1 for the anchored departments and Bogot\'a. For the nine partially represented territories the factor is the share of the capital, drawn as straight segments between January 2001, January 2014 and December of the last anchored year, with later months following the last slope; a factor that follows the monthly noise of two independent reconstructions would transmit that noise to the represented universe. For 1993--2000 there is no national GEIH coverage to refer to: anchored departments and Bogot\'a carry factor 1 as a conservation convention, the nine territories carry their capital share, and both cases are flagged as assumed.

\subsubsection{History before 2001}

Months before January 2001 are not reconciled to a national series, because none was published. From 2001 the national population of the survey agrees with the projections of the territories it represents within 0.1\%; before 2001 the reconstructed national reference falls up to 1.6\% below them. For those months every territory is placed on its own demographic projection and its labor components are rescaled by the same factor, which preserves every rate exactly; the ICML national total is their sum.

\subsection{Reconstruction of 1993--2000}
\label{sec:reconstruction}

The months before the first GEIH month are a reconstruction and two properties are imposed on them, for the national reference, every department and every city, each with its own official months.

\subsubsection{Linear working-age population}

The working-age population is a demographic stock and cannot oscillate. It is drawn as a straight line from the median of its first year to its value in January 2001,
\begin{equation}
\PET_{t} = \PET_{t_0} + \frac{t - t_0}{t_1 - t_0}\left(\PET_{t_1} - \PET_{t_0}\right), \qquad t_0 \le t \le t_1,
\label{eq:linear-pet}
\end{equation}
with $t_1$ = January 2001, so that a single slope runs through the seam. Total population before 2001 is drawn in the same way. The annual means of the ENH working-age population are not collinear (the survey was redesigned in 1996) and are therefore no longer reproduced.

\subsubsection{Generated dynamics}

The levels are rebuilt on the linear working-age population keeping the reconstructed rates, $E_t = \TO_t\,\PET_t$ and $\PEA_t = E_t/(1 - \TD_t)$, and the rates receive the short-run dynamics of the observed period:
\begin{enumerate}
  \item The irregular component of the official $\TO$ and $\TD$ (from 2001) is extracted with a robust STL decomposition of period 12 and seasonal window 13.
  \item Deterministic noise is added to $\TO$ and $\TD$ at a scale of 0.45 times the standard deviation of that irregular component. The pseudo-random seed of a month is a base seed plus a hash of the month label, plus the territory code, so a month always receives the same draw regardless of processing order and territories do not share draws.
  \item The seasonal component of employed and unemployed is replaced by the 2001--2018 seasonal pattern of the same series, scaled by the ratio between the mean of the history and the mean of the reference period; the STL trend and irregular components of the history are kept.
  \item The accounting closure is exact: $\PEA_t = E_t + U_t$ and $I_t = \PET_t - \PEA_t$.
\end{enumerate}
The generated noise and seasonality give the history the statistical behavior of the survey period. They are a continuity device and carry no information about what happened in a particular month between 1993 and 2000.

\subsubsection{Smoothing of territorial denominators after 2001}

After January 2001 the national series are official and are not modified, but the departmental working-age population produced by reconciliation was serrated (a median of 153 changes of direction in 308 months, against none in the national series). The \emph{share} of each department in the national total is smoothed with the minimum-curvature operator
\begin{equation}
\hat{\bm{z}} = \argmin_{\bm{z}} \; \|\bm{z} - \bm{x}\|_2^{2} + \lambda\,\|\bm{D}_2 \bm{z}\|_2^{2}
\quad \text{s.t. } \bm{A}\bm{z} = \bm{A}\bm{x},
\label{eq:min-curvature}
\end{equation}
where $\bm{D}_2$ is the second-difference operator and $\bm{A}$ averages months into calendar years. The operator is linear and maps a constant to itself, so the 33 smoothed shares still add to one and the departments still add exactly to the ICML national total. Employed, unemployed and labor force are unchanged; the population outside the labor force absorbs the change, and the employment and participation rates are recomputed on the new denominator. The unemployment rate does not change. Cities, which add to nothing, are smoothed in levels with the same operator.

\subsection{Quality gate and labor products}
\label{sec:quality}

The quality stage does not re-estimate anything. It verifies and exports three tables: cities (33 units per month), departments (33 units per month) and nationals (two rows per month, the ICML total and the DANE reference). Every table covers every month from January 1993 to the cutoff with no missing level. The stage blocks on any violation of \eqref{eq:id-pea}--\eqref{eq:id-pop} above $10^{-5}$ persons, on duplicated or missing keys, on negative components, and on a difference with the national reference above the rounding of the source. It also computes non-blocking diagnostics: changes by a factor of three or more, sustained shifts in rates over 12-month windows (thresholds of 5 percentage points for unemployment and 4 for employment and participation), demographic jumps above 1\% per month, and the linearity of each territory-year.

\section{Informality}
\label{sec:informality}

No official source observes informal employment by department. The national and urban figures that exist are published for moving quarters and under two operational definitions. The informality products are therefore the most model-dependent part of the system, and the chain is designed so that the only quantities forced on the departments are a national target and their own employment.

\subsection{Evidence regimes}

Table~\ref{tab:informality-regimes} lists the regimes. Each one is a different estimand, and the labels are carried to every record of the products as a provenance and an uncertainty class.

\begin{table}[H]
\centering
\caption{Evidence regimes of informality.}
\label{tab:informality-regimes}
\footnotesize
\begin{tabularx}{\textwidth}{L{2.3cm} L{5.0cm} Y}
\toprule
\textbf{Period} & \textbf{Source} & \textbf{Treatment} \\
\midrule
1993--2000 & No informality observation; labor reconstruction only & Backcast of the national rate without an official anchor; uncertainty class \emph{very high} \\
2001--2006 & 13 metropolitan areas aggregated; criterion of establishments of up to five workers; April--June windows & Six aggregate anchors, transferred to the rest of the country; class \emph{high} \\
2007--March 2021 & 23 cities and 13 areas; same criterion; 3-month moving windows & Regularized inversion of rates into latent months; class \emph{high, definition bridge} \\
2020 & Publication interrupted & Three scenarios for the gap; one adopted and two kept as sensitivity \\
2021--cutoff & Current definition; 23 cities, 13 areas and national total; 3-month moving windows & Inversion of counts into latent months; class \emph{modeled, current definition} \\
\bottomrule
\end{tabularx}
\end{table}

\subsection{National latent series}

A published moving-quarter mean constrains three consecutive months. With $K$ windows there are $K+2$ months and the system has a two-dimensional null space. Let $\bm{x}_0$ be a particular solution and $\bm{Z}$ a basis of the null space of the window operator. The latent monthly path is the member of the solution set with minimum second difference,
\begin{equation}
\hat{\bm{x}} = \bm{x}_0 + \bm{Z}\hat{\bm{\theta}}, \qquad
\hat{\bm{\theta}} = \argmin_{\bm{\theta}} \left\| \bm{D}_2\left(\bm{x}_0 + \bm{Z}\bm{\theta}\right) \right\|_2^{2},
\label{eq:rolling3}
\end{equation}
which reproduces every published window within $10^{-4}$ persons. For the earlier regime, where rates rather than counts are published and the windows are noisier, the inversion is regularized instead of exact. The 2020 gap is filled with an interpolated scenario; a scenario that repeats the 2019 seasonal profile and one that transmits the urban shock are computed for the twelve months and published as alternatives.

The 2021 change of definition is treated as a break between regimes: the series are joined with a documented bridge and are never concatenated as if the definitions were equal.

\subsection{Urban sequence}

The urban reconstruction follows the order in which evidence exists: the 13-area aggregate, then the national series, then the 23 cities with published windows, then the 33 urban signals. Cities without their own windows use the donors selected in the labor chain (Section~\ref{sec:urban}) and Cundinamarca uses its synthetic signal. As in the labor products, urban informality signals are non-additive. The sequence is regenerated from the annexes and the current labor product in every run and compared with the frozen sequence before it is used.

\subsection{Departmental priors from a residual network}
\label{sec:mlp}

Departmental rates are not observed, so a model is trained where labels exist (cities) and applied where they do not (departments). The labels are the latent monthly informality rates of the 23 cities with published windows, from 2007 to 2025 excluding 2020 (4,968 city-months). The predictors are 38 labor-market features available for both cities and departments: levels and rates of the labor quantities and their cluster-relative versions (cluster means, residuals and ratios), with the territorial clustering policy of Section~\ref{sec:modeling} refitted on training cities in each fold.

The network maps the standardized feature vector $\bm{x} \in \mathbb{R}^{p}$ to a logit:
\begin{align}
\bm{h}_0 &= \mathrm{Dropout}\left(\mathrm{ReLU}(\bm{W}_0\bm{x} + \bm{b}_0)\right), & \bm{h}_0 &\in \mathbb{R}^{128}, \nonumber\\
\bm{h}_1 &= \bm{h}_0 + \mathrm{ReLU}\left(\mathrm{Dropout}\left(\bm{W}_1\,\mathrm{LayerNorm}(\bm{h}_0) + \bm{b}_1\right)\right), \nonumber\\
\eta &= \bm{w}_2^{\top}\bm{h}_1 + b_2, & \mu &= \expit(\eta),
\label{eq:mlp}
\end{align}
with one residual block and dropout probability 0.1. The loss is the negative log-likelihood of a Beta distribution with mean $\mu$ and fixed precision $\phi = 20$ \citep{FerrariCribariNeto2004}:
\begin{equation}
\ell(\eta; y) = \log \mathrm{B}\left(\mu\phi, (1-\mu)\phi\right) - (\mu\phi - 1)\log y - \left((1-\mu)\phi - 1\right)\log(1 - y),
\label{eq:beta-nll}
\end{equation}
including the normalizing term. The Beta density is a working likelihood for rates in $(0,1)$ and not a claim that latent monthly rates are independent Beta draws. Training uses the Adam optimizer with learning rate $10^{-3}$, weight decay $10^{-3}$, batches of 512, at most 160 epochs and early stopping with a patience of 18 epochs on the validation split.

The model is trained once. Its output for the 33 departments and every month is stored as a frozen parameter, and the network is not retrained in production (Section~\ref{sec:frozen}). The priors are a statement about the \emph{pattern} across departments; their level is set by the calibration below.

\subsection{Monthly closure}
\label{sec:h9}

For each month, let $p_d \in (0,1)$ be the prior rate of department $d$, $E_d$ its employment in the census universe, and $F^{\mathrm{inf}}_{\mathrm{nat}}$ the national target of informal employment in that universe (the latent national rate times the ICML national employment). The calibrated rates are those closest to the priors in Kullback--Leibler divergence between Bernoulli distributions, weighted by employment, subject to the national count:
\begin{equation}
\min_{\{q_d\}} \sum_{d} E_d \left[ q_d \log\frac{q_d}{p_d} + (1 - q_d)\log\frac{1 - q_d}{1 - p_d} \right]
\quad \text{s.t. } \sum_d E_d\, q_d = F^{\mathrm{inf}}_{\mathrm{nat}}.
\label{eq:kl-problem}
\end{equation}
The solution is a common shift on the logit scale,
\begin{equation}
q_d = \expit\left(\logit(p_d) + \gamma\right),
\label{eq:kl-solution}
\end{equation}
with $\gamma$ the unique root of $\sum_d E_d\,\expit(\logit(p_d) + \gamma) = F^{\mathrm{inf}}_{\mathrm{nat}}$, found by bisection to a tolerance of $10^{-4}$ persons. This is a calibration in the sense of \citet{DevilleSarndal1992}: the ordering of departments given by the priors is preserved and only the level moves. Informal and formal employment follow as $F^{\mathrm{inf}}_{d} = q_d E_d$ and $F^{\mathrm{for}}_{d} = E_d - F^{\mathrm{inf}}_{d}$, so identity \eqref{eq:id-inf} holds by construction.

Four alternative calibrators were compared (the Bernoulli divergence above, an employment-weighted quadratic distance, a variance-scaled quadratic distance and a quadratic distance with a temporal penalty on the correction). Their mean interventions coincide within 0.001 percentage points, and the divergence-based calibrator was kept.

Only the census universe is calibrated to a national margin. The represented universe is derived by applying the calibrated rate to the represented employment, $F^{\mathrm{inf,rep}}_{d} = q_d\,E^{\mathrm{rep}}_{d}$, which assumes that the informality rate of the unrepresented part of a territory equals that of its represented part. That assumption is recorded as a flag in every record. Closing the represented universe first and deriving the census universe instead moves the other national margin by 0.19--0.21 percentage points; the census-first direction was adopted.

The link between the whole territory and its represented part for the nine partially represented territories is anchored to the adjusted-coverage population of the municipal head of each capital in the 2018 census. The bridge is a frozen parameter: its levels are refreshed from the current labor product whenever they differ by more than $10^{-3}$ persons, and its factors are those of the labor product (Section~\ref{sec:representativeness}), so that labor and informality describe the same universe.

City informality applies the urban latent rate to the city employment of the labor product. The national series of both universes are the sums of the departmental counts.

\subsection{Splice at January 2001}
\label{sec:splice}

Until December 2000 the informality rate is a backcast with no official anchor; from January 2001 it is anchored to published figures. The two segments do not meet: before treatment, the national rate rises 2.78 percentage points in one month, about three times the 99th percentile of its monthly changes, and the step is present in the 33 cities and in 29 of the 33 departments. It is a difference of level between two measurement regimes.

The break is estimated by a regression with a jump and a common slope over a window of $h = 24$ months on each side of the break month $t^\ast$:
\begin{equation}
r_t = a + b\,(t - t^\ast) + \delta\,\mathbb{1}\{t \ge t^\ast\} + \varepsilon_t, \qquad |t - t^\ast| \le h.
\label{eq:rd}
\end{equation}
A difference of annual means would mix the break with a year of drift, and a one-month difference would amplify noise. The side with information stays and the side without it moves: the segment before $t^\ast$ is shifted by the constant $\hat{\delta}$, with its monthly dynamics intact. The shift is applied to the national target, to every city and to the priors of every department, which are then recalibrated with \eqref{eq:kl-solution} so that they still add to the national target. Everything from January 2001 onward is unchanged.

\section{Employment by Sector}
\label{sec:sectoral}

The sectoral product distributes the employment of each territory-month among 15 groups of economic activity (Appendix~\ref{app:sectors}). It is \emph{experimental}: departmental sector labels are never observed monthly, the territorial structure before 2015 rests on a declared assumption, and the product never replaces or modifies the labor bases. It is included because it is reconciled to them and audited with them.

\subsection{Observed support}

\begin{table}[H]
\centering
\caption{Observed support of sectoral employment.}
\label{tab:sector-support}
\footnotesize
\begin{tabularx}{\textwidth}{L{2.2cm} L{5.2cm} Y}
\toprule
\textbf{Period} & \textbf{Source} & \textbf{What exists} \\
\midrule
2015--cutoff & Monthly GEIH annex & Thirteen published activity groups per calendar month, national only. Mining and utilities are published together; their split is an assumption with a fixed ratio \\
2007--2014 & Historical annexes under the previous classification, moving quarters & Normalized window observations; no monthly observation by group \\
2001--2015 & Bridge between classifications on the 2015 overlap & National shares only, under three traceable scenarios \\
1993--2001 & ENH, four broad groups, September & Homologation to the 15 groups; exploratory \\
2015--2025 & Annual sectoral shares for 23 departments, plus an external reference for Bogot\'a & Annual anchors, used as soft priors \\
\bottomrule
\end{tabularx}
\end{table}

Two rules hold throughout. A moving quarter or an annual figure constrains its own support and is never promoted to a calendar month. Observed figures are never rewritten.

\subsection{Chain}

\subsubsection{National monthly composition}

A share per sector and month, $\pi_{s,t}$ with $\sum_s \pi_{s,t} = 1$, is assembled from January 1993 to the cutoff. From 2015 the calendar-month shares are read from the annex. Between 2001 and 2015 the moving-quarter observations are inverted into months with a non-negative, source-aware inverse that trades fidelity against identification, since exact reproduction of three-month windows leaves a two-dimensional null space. Before 2001 the four published broad groups constrain broad trajectories and the shares of the modern subgroups within each broad group are held at their first linked proportions. Every record carries a status that states what is observed and what is reconstructed.

\subsubsection{Departmental priors}

Composition priors for each department are built from population, the departmental accounts by activity and the national composition. They are explicit priors and not estimated regression coefficients.

\subsubsection{Territorial predictor}

For the years with annual departmental shares (2015--2024 for training, 2025 held out), a predictor of the 14 productive shares is fitted on 38 labor-market features of the department and lagged departmental GDP by activity. The predictor is an equally weighted ensemble of a ridge regression and gradient-boosted trees, with the weight chosen from a predefined grid $\{0, 0.25, 0.5, 0.75, 1\}$ by leave-one-department-out validation on 2015--2024. Bogot\'a never enters training; its external reference is used only for evaluation and as a soft anchor.

\subsubsection{Reconciliation}

For each month, the prior table $\tilde{P}_{d,s}$ of departmental sector employment is adjusted by iterative proportional fitting \citep{DemingStephan1940} so that rows add to the employment of each department and columns to the national employment of each sector:
\begin{equation}
\min_{\bm{P} \ge 0} \sum_{d,s} P_{d,s}\log\frac{P_{d,s}}{\tilde{P}_{d,s}}
\quad \text{s.t. } \sum_{s} P_{d,s} = E_{d,t}, \quad \sum_{d} P_{d,s} = \pi_{s,t}\,E^{\mathrm{ICML}}_{t}.
\label{eq:sector-raking}
\end{equation}
The annual departmental shares enter as soft priors and are not imposed exactly, because they refer to the represented universe and are not compatible with exact national margins of the census universe. For the represented universe the same shares are applied to the represented employment of each department.

\subsubsection{Refit in production}

Unlike the informality priors, the sectoral predictor is refitted in every recomputation with the current labor base, from frozen inputs, so that the sectoral product is always consistent with the labor product it distributes.

\section{Production System: Vintages, Gates and Reproducibility}
\label{sec:system}

A reconstructed panel is defined by its method and by the rules that decide when a published figure may change. This section describes those rules as implemented in version 2.6.1 of the software \citep{ICML_Repository}.

\subsection{Inputs and the two-annex rule}
\label{sec:two-annex}

A month $M$ is processed only when two official publications exist for it: the monthly labor-market annex whose reference period is $M$, and the informality annex whose moving quarter ends in $M$ (it must span exactly $M-2$, $M-1$ and $M$). If either is missing, cannot be verified or declares another period inside the workbook, no estimate of any product is generated for $M$: the last available informality rate is not carried forward and annexes of different months are not combined. The cutoff of a run is the last month with both annexes.

The file name classifies an annex and its content decides: a workbook whose internal period differs from its name blocks the run, as do two different files that declare the same month.

\subsection{Frozen parameters}
\label{sec:frozen}

Three sets of parameters are estimated once and reused without retraining in ordinary updates:
\begin{itemize}
  \item the bridge between universes anchored to the 2018 census;
  \item the departmental informality priors of the residual network (Section~\ref{sec:mlp});
  \item the input features of the sectoral predictor.
\end{itemize}
The reasons are comparability (retraining every month would change the history for reasons unrelated to the new data), traceability, and independence of an ordinary update from the deep-learning stack. A change of source or method (a redesign of the survey, new population projections, a new census, a change of definition) is not an ordinary update: it requires full reprocessing and a new version, and earlier vintages are kept.

The frozen panels end at the cutoff of the reference vintage (August 2026). For a later month they are extended with explicit rules, and every extended record carries a tag that reaches the published products:
\begin{itemize}
  \item informality priors: the same calendar month of the previous year (the priors set the departmental pattern; the level comes from the national target of the new annex);
  \item bridge between universes: the last frozen month for the factors of the fully covered territories, while levels are refreshed from the labor product and the nine partial territories take the labor product's factors;
  \item national sectoral composition: read from the labor annex of the new month when it publishes it, and otherwise the same month of the previous year.
\end{itemize}
The rules are covered by unit tests and by a test on extended panels. An end-to-end run with real annexes of a month after August 2026 has not been executed at the time of writing.

\subsection{Revision states and updates that do not rewrite history}
\label{sec:update}

Each cell carries a revision state that depends only on dates: the last month of the vintage, the last year with a published departmental annex, and the 12-month window of the city publications.

\begin{table}[H]
\centering
\caption{Revision states.}
\label{tab:states}
\small
\begin{tabularx}{\textwidth}{L{3.0cm} Y L{4.2cm}}
\toprule
\textbf{State} & \textbf{Meaning} & \textbf{Changes in an ordinary update} \\
\midrule
Final & Its year has a published annual anchor and no future window touches it & No \\
Subject to anchor & Its windows are closed but its year has no published anchor yet & When the anchor is published \\
Provisional & Inside the open 12-month moving window & Every month \\
\bottomrule
\end{tabularx}
\end{table}

A vintage with last month $M$ can still revise months from $M-10$ onward, because the window ending in $M+1$ covers them. Recomputing the whole chain when a month is added would change closed history: in a test with two consecutive vintages built from the same annex (54 and 57 windows), a full recomputation changed 16,932 final cells, by up to 2\% in unemployed, because every new window re-identifies the latent months from 2021 onward.

Two modes follow. In the \textbf{anchored} mode (the default), everything is recomputed and then the years covered by the departmental annex are grafted from the published vintage. Whole rows are taken, never single cells, so each row keeps the identities, factors and national closure of one computation. The run is refused if any final cell changes, naming the cells. In the two-vintage test the graft changed no final cell, changed 1,353 provisional cells and added 99 rows (3 months for 33 territories), with identities exact to $10^{-9}$ persons. In the \textbf{total} mode, the whole series is rewritten; it requires explicit confirmation and is reserved for methodological changes. A revision by the source of a figure of a closed year is detected by comparing consecutive releases of the annex through the same readers the chain uses, and blocks an anchored run.

\subsection{Gates}
\label{sec:gates}

A result is accepted only if it passes every blocking gate (Table~\ref{tab:gates}). Diagnostic checks and scientific-evidence checks are recorded and do not stop a run. In particular, the informality quality gate contains four evidence checks that fail by design (the definition bridge before 2021 is not independently identified; 2020 has no monthly observations; the informality of the unrepresented rural complement is not observed; the represented national rate is not independent of the census-universe rate), and the system never authorizes publication by itself.

\begin{table}[H]
\centering
\caption{Blocking gates and their limits.}
\label{tab:gates}
\footnotesize
\begin{tabularx}{\textwidth}{L{3.1cm} Y L{3.6cm}}
\toprule
\textbf{Moment} & \textbf{Control} & \textbf{Limit} \\
\midrule
Reception & Pair of annexes of the same month; content period equal to the name's; no ambiguous files & Exact \\
Anchored plan & Fingerprint of the frozen parameters unchanged; modified inputs only with explicit authorization & Exact \\
Preparation & All extractors finish correctly & 9 of 9 \\
National stage & Annual and monthly identities; relative displacement of the reconstruction & 1 person; 10\% annual, 15\% monthly \\
Reconciliation & National, joint and anchor closure & $10^{-3}$ persons \\
Labor quality & Accounting identities of the three tables & $10^{-5}$ persons \\
Informality & Regenerated sequence equivalent to the frozen one; completeness; identities & Comparison tolerances \\
Reconstruction and splice & Identities exact; working-age population linear in 1993--2000; official months identical; shares in $(0,1]$ & $10^{-6}$ persons; $10^{-9}$ in shares \\
Cross audit & Coherence between labor, informality and sectoral products & $10^{-3}$ persons; shares $10^{-8}$ \\
Reference equivalence & At the frozen cutoff, the eight candidates reproduce the frozen products & 0.1 persons; rates and factors $10^{-6}$ \\
Anchored mode & No cell of a closed year changes after the graft & Exact \\
Delivery & 52 checks: hashes, shapes, identities between universes, full coverage & 52 of 52 \\
\bottomrule
\end{tabularx}
\end{table}

\subsection{Reference vintage and numerical reproducibility}
\label{sec:reference}

The eight products at the August 2026 cutoff are stored in the repository with their hashes as the \emph{reference vintage}. A recomputation at that cutoff must reproduce them; a mismatch ends the run in a failure state that requires an explicit refreeze. This turns ``the numbers did not change'' into a test: a change of layout, naming or performance must leave the products byte-identical, and a methodological change must be accompanied by a refreeze, a new version and a changelog entry with the metrics that moved.

The reference was measured against recomputations under different dependency versions:
\begin{itemize}
  \item in the locked environment, the eight products are byte-identical;
  \item with the latest releases of the numerical libraries and the same minor release of the boosting library, differences are at most 0.01 persons and $10^{-9}$ in rates, which is floating-point noise and is accepted by the gate;
  \item with a later minor release of the boosting library, the estimates that depend on the modeling stage change by up to 38,900 persons of departmental working-age population, 1,050 persons of national labor force and 1.3 percentage points of a sectoral share.
\end{itemize}
The boosting library is therefore capped to one minor series, and raising the cap is treated as a methodological change. Tree models are also not bit-reproducible across operating systems and processor builds. The platform on which the reference was frozen is recorded next to it; on another platform the comparison accepts a bounded, purely numerical difference (no cell may move by more than 12\% of the typical magnitude of its column, no key or category may differ and the cross audit must pass), reports the measured maximum, and labels the run accordingly. Any computer can become the strict reference by recomputing and refreezing.

\subsection{Deliveries}
\label{sec:delivery}

A delivery is written only after the recomputation ends in a clean state and the cross audit passes. Each product is split by universe according to a declarative specification: the census universe receives the whole-territory levels; the GEIH universe receives the represented levels, the rates derived from them and the coverage status of each department-month; the national labor figure of the GEIH universe is the one published by DANE, with the sum of the represented departments reported as a diagnostic; and cities are the same series in both universes. The delivery contains 16 files (eight products in two universes), a manifest with the hash of each file, the column map, the software version, the input fingerprints and the active methods, and the report of the 52 checks. The checks re-derive relations that must hold in the data (complete level times factor equals represented level; complete equals represented plus outside; sector totals equal labor employment), so an incorrect column assignment fails even if the mechanical split succeeded. Output folders must be new: the system refuses to overwrite a delivery. A simplified version without traceability columns is cut from the verified delivery.

\subsection{Software}

The system is a single Python package installable from the standard package index, with one command-line interface. Listing~\ref{lst:cli} shows the routine use and Listing~\ref{lst:config} an excerpt of a stage configuration; Listing~\ref{lst:read} shows how the labor tables are read.

\begin{lstlisting}[language=bash, caption={Rebuild through a month, update with the next one, verify and report.}, label={lst:cli}]
icml status                          # inputs, published base, next step
icml rebuild --through 2026-07 --yes # recompute the history from 1993
icml update                          # add the processable months (anchored)
icml verify                          # run the 52 delivery checks again
icml report                          # diagnostics of the newest month
icml checksums                       # verify the checkout against SHA256SUMS
\end{lstlisting}

\begin{lstlisting}[language=toml, caption={Excerpt of the reconstruction configuration of the reference vintage.}, label={lst:config}]
[labor_reconstruction]
enabled = true
splice_month = "2001-01"
scaling_factor = 0.45
base_seed = 42
stl_period = 12
stl_seasonal = 13
stl_robust = true
reference_first_year = 2001
reference_last_year = 2018
linearize_population = true
smooth_territories = true
\end{lstlisting}

\begin{lstlisting}[language=Python, caption={Reading the three labor tables. Territory codes are text; the two national series are never added.}, label={lst:read}]
import pandas as pd
from pathlib import Path

p = Path("quality/products")
cities = pd.read_csv(p / "ciudades.csv",
                     dtype={"city_code": "string", "dep_code": "string"})
departments = pd.read_csv(p / "departamentos.csv", dtype={"dep_code": "string"})
nationals = pd.read_csv(p / "nacionales.csv", dtype={"dep_code": "string"})

icml_total = nationals.query("national_type == 'ICML_COMPLETO'")
dane_reference = nationals.query("national_type == 'DANE_REFERENCIA'")

# The ICML national total is the sum of the 33 whole-territory levels.
check = departments.groupby("period")["employed"].sum()
\end{lstlisting}

Every monthly delivery is accompanied by a diagnostic report of the newest month: for each monitored series, the month-on-month change and its $z$-score against the distribution of such changes over the previous ten years (an absolute value above 3 is flagged), the year-on-year change, the residual with respect to a seasonal-naive prediction, and the revisions with respect to the previous delivery. Univariate 12-month forecasts are available as an experimental tool and are not part of the products.

\section{Results}
\label{sec:results}

All figures and tables in this section are computed from the reference vintage (cutoff August 2026, software version 2.6.1), which is distributed with the repository. The panel has 404 months, 13,332 territory-months in the departmental tables and 808 rows in the national tables. Validation evidence is presented first, followed by the resulting series.

\subsection{Deviation from official figures}
\label{sec:deviation}

For every territory, year and quantity with an official figure, the published estimate is compared with that figure: the 23 anchored departments against the annual departmental annex (1,748 comparisons), the nine capitals against their published 12-month windows (468) and Bogot\'a against its urban domain (104), a total of 2,320 comparisons from 2001.

\begin{table}[H]
\centering
\caption{Absolute deviation of published estimates from the official figure they correspond to (2,320 territory-year-quantity comparisons).}
\label{tab:deviation}
\small
\begin{tabular}{lrrr}
\toprule
\textbf{Quantity} & \textbf{Median (\%)} & \textbf{95th percentile (\%)} & \textbf{Maximum (\%)} \\
\midrule
Population & 0.04 & 0.14 & 0.33 \\
Employed & 0.26 & 0.98 & 1.22 \\
Outside the labor force & 0.30 & 1.66 & 1.66 \\
Unemployed & 0.43 & 1.86 & 1.89 \\
\bottomrule
\end{tabular}
\end{table}

The medians are similar across the three types of reference (0.2--0.3\% in employed, 0.3--0.5\% in unemployed), so no group of territories carries the burden. The maxima are concentrated in 2020 and 2021, the years of the pandemic and of the change of sampling frame; outside those years every deviation is below 0.8\%. This is the deviation of the whole method (sources, denominators, disaggregation, latent months and closure) and not the error of a predictive model. It has two components: the share of the sources' own non-additivity that each territory absorbs, which is published per cell and is therefore separable, and the remainder, which is attributable to the construction.

\subsection{Who absorbs the non-additivity of the sources}
\label{sec:residual-evidence}

Table~\ref{tab:residual} compares the two allocation rules of Section~\ref{sec:representativeness}. Keeping every annual anchor exact concentrates the inconsistency of the sources on the units without an annual anchor, which are themselves official survey domains: Bogot\'a deviates by up to 9.1\% from its published urban figures and a capital by up to 12.1\% from its published windows. The proportional rule keeps every official figure within 1.9\%.

\begin{table}[H]
\centering
\caption{Maximum deviation from official figures under the two rules for allocating the national residual.}
\label{tab:residual}
\small
\begin{tabular}{lrr}
\toprule
\textbf{Measure} & \textbf{Anchors exact} & \textbf{Proportional (adopted)} \\
\midrule
Annual anchor deviation, employed & 0.00\% & 0.98\% \\
Annual anchor deviation, unemployed & 0.00\% & 1.86\% \\
Annual anchor deviation, outside the labor force & 0.00\% & 1.66\% \\
Annual anchor deviation, population & 0.00\% & 0.14\% \\
Bogot\'a against its official urban domain & 9.08\% & 1.86\% \\
A capital against its official windows, maximum & 12.10\% & 2.80\% \\
Capitals, 90th percentile & 3.50\% & 0.80\% \\
\bottomrule
\end{tabular}
\end{table}

The cost of the adopted rule is explicit: annual departmental anchors are reproduced within a declared and tested bound and no longer exactly. At the monthly frequency the deviation has a median of 0.3\% in employed and 0.5\% in unemployed, with 95th percentiles of 0.8\% and 2.5\%.

\subsection{Transfer evaluation: treating an observed department as unobserved}
\label{sec:transfer}

The accuracy of the estimates for territories without official figures cannot be measured directly. The closest measurable analogue is to hide a territory that has official figures and estimate it as if it had none. Each of the 22 anchored non-synthetic departments is hidden in turn: models and clusters are refitted without it, the panel is built without its references and the closure runs without its anchors. Its annual means for 2007--2025 are then compared with its official anchors (418 department-years per candidate), before and after the hierarchical closure. A leakage control verifies that multiplying all labels and anchors of the hidden department by 1.5 leaves its estimates unchanged.

\begin{table}[H]
\centering
\caption{Leave-one-department-out evaluation: mean absolute error over 418 department-years. Rates in percentage points; levels in per cent. ``Model'' is the prediction before closure; ``Closed'' applies the closure that keeps annual anchors exact.}
\label{tab:transfer}
\footnotesize
\begin{tabular}{llrrrrrrr}
\toprule
\textbf{Candidate} & \textbf{Stage} & \textbf{Rates} & $\TD$ & $\TO$ & $\TGP$ & \textbf{Unemp.} & \textbf{Work.\ age} & \textbf{Pop.} \\
\midrule
Hybrid portfolio & Closed & 2.21 & 1.43 & 2.52 & 2.67 & 16.1 & 5.27 & 2.67 \\
Urban analog & Closed & 2.23 & 1.49 & 2.63 & 2.57 & 18.7 & 5.23 & 2.67 \\
Boosted trees, Ward clusters & Closed & 2.90 & 1.42 & 3.43 & 3.86 & 17.8 & 5.27 & 2.67 \\
Boosted trees, no clusters & Closed & 2.93 & 1.47 & 3.44 & 3.86 & 18.5 & 5.61 & 2.67 \\
National rates & Closed & 5.14 & 1.89 & 6.57 & 6.96 & 23.1 & 5.23 & 2.67 \\
\midrule
Boosted trees, Ward clusters & Model & 2.34 & 1.52 & 2.63 & 2.88 & 17.1 & 5.17 & 0.59 \\
Boosted trees, no clusters & Model & 2.34 & 1.53 & 2.63 & 2.85 & 17.4 & 5.44 & 0.59 \\
Hybrid portfolio & Model & 2.37 & 1.52 & 2.68 & 2.92 & 17.2 & 5.17 & 0.59 \\
Urban analog & Model & 2.72 & 2.45 & 2.78 & 2.95 & 30.3 & 5.34 & 0.59 \\
National rates & Model & 3.51 & 2.01 & 4.22 & 4.29 & 22.6 & 5.34 & 0.59 \\
\bottomrule
\end{tabular}
\end{table}

Four findings follow from Table~\ref{tab:transfer} and from paired differences by department with 10,000 bootstrap replicates.

\begin{enumerate}
  \item The ranking of candidates reverses after the closure. Before it, the boosted model with Ward clusters is the best candidate; after it, it is 0.70 percentage points worse than the hybrid portfolio (95\% interval 0.37 to 1.02; worse in 18 of 22 departments). A model evaluated only before reconciliation is not evaluated in the form in which it is published.
  \item After the closure, the departmental model adds no measurable value over the urban signal. The difference between the hybrid portfolio and the urban analog is $-0.03$ points (interval $-0.15$ to $0.09$). The urban signal carries most of the information: it improves on national rates by 2.9 points.
  \item A closure that forces the departments into the national survey total degrades population. The demographic projection misses the official departmental population by 0.59\%; the population reconciled to the survey total misses it by 2.67\%, and by 7.4\% and 6.8\% when a large department (Antioquia, Valle del Cauca) is hidden, because the coverage gap of the national total is forced onto the unanchored subset.
  \item The working-age population is the largest level error (about 5\% for every candidate, before and after the closure): the denominator, not the rates, dominates the errors in levels.
\end{enumerate}

These findings motivate the published design. Since the gap between universes is survey coverage (Section~\ref{sec:sources}), departments are published for the whole territory with explicit factors instead of being forced into the national total; and since the model does not improve on the survey's own urban domains after closure, the territories without an annual anchor are built from those domains (Section~\ref{sec:representativeness}) and the model is confined to provisional months.

Three alternatives to the closure that keeps anchors exact were evaluated under the same protocol. A uniform raking of all 33 territories moves official anchors by up to 3.1\% (employed) and 7.1\% (population) and was rejected. Reconciling components only, without population, removes the population error at no cost in rates but leaves the working-age population in the narrower universe and the population in the wider one. Scaling the national total to the demographic universe keeps anchors exact, improves population in 22 of 22 departments (mean absolute error 0.58\% instead of 2.67\%) and yields the lowest rate error with the hybrid portfolio (2.15 points), at the explicit cost that the departmental product adds to a universe 1--5\% larger than the published national total. The published design adopts that logic and makes the difference explicit through the factors.

Anchored departments are larger and better covered by urban signals than the territories without an anchor. The evaluation is evidence of transferability and not an error estimate for those territories. Hyperparameters and architecture were chosen with all departments, the predictors use the August 2026 vintage rather than the vintage available at each date, and only 2007--2025 is evaluated.

\subsection{Rest of territory of the partially represented departments}
\label{sec:rest-evidence}

When a represented department is hidden, the closure can recover part of it from the national residual. The rest of the nine partially represented territories is not inside the national total, so that mechanism does not apply to them. The appropriate analogue is to predict the rest of territory (department minus capital) of the 22 anchored non-synthetic departments, where it is observable on annual means.

\begin{table}[H]
\centering
\caption{Mean absolute error (percentage points) of alternative rules for the labor ratios of the rest of territory, evaluated on anchored departments.}
\label{tab:rest}
\small
\begin{tabular}{lrrrr}
\toprule
\textbf{Rest ratios taken from} & \textbf{Rates} & $\TD$ & $\TO$ & $\TGP$ \\
\midrule
Capital participation and mean unemployment rate of others & 2.41 & 1.42 & 2.88 & 2.95 \\
Mean of the other departments' rest (adopted) & 2.48 & 1.39 & 2.96 & 3.08 \\
Pooled rest of the other departments & 2.51 & 1.41 & 3.01 & 3.10 \\
Own capital & 2.72 & 2.45 & 2.78 & 2.95 \\
Most similar department & 2.96 & 1.97 & 3.32 & 3.60 \\
\bottomrule
\end{tabular}
\end{table}

The first three rules are statistically indistinguishable (for instance, the difference between the mean and the pooled rule is $-0.03$ points, interval $-0.09$ to $0.03$). The mean is adopted because it is a single rule for all ratios and does not use the participation of the capitals, which for the three smallest ones rests on poorly conditioned latent months. Errors grow as the population share of the capital falls: for departments whose capital holds 15--19\% of the population the error is 4.5--7.4 points. Among the nine territories, Putumayo (0.11) and Vichada (0.13) are in that regime. The donors' rest is Andean and Caribbean rural territory; no survey represents the Amazonian rest.

\subsection{Properties of the reconstructed history and of the denominators}

Table~\ref{tab:reconstruction} reports the properties that Section~\ref{sec:reconstruction} imposes, measured on the reference vintage.

\begin{table}[H]
\centering
\caption{Measured properties of the 1993--2000 reconstruction and of the territorial denominators.}
\label{tab:reconstruction}
\footnotesize
\begin{tabularx}{\textwidth}{Y R{3.4cm}}
\toprule
\textbf{Measure} & \textbf{Value} \\
\midrule
Deviation of the national working-age population from a straight line, 1993--1999 & 0 \\
Change of the working-age population from December 2000 to January 2001 & $+0.28\%$ (one month of slope) \\
Maximum accounting residual, labor tables & below $10^{-8}$ persons \\
Sum of the 33 departments minus the ICML national total, maximum & below $10^{-8}$ persons \\
Mean absolute monthly change of the national unemployment rate, ICML total, 1993--1999 / 2001--2010 & 0.80 / 0.91 points \\
Same measure, DANE reference & 0.95 / 0.92 points \\
Departmental working-age population after 2001, changes of direction (median of 33), before / after smoothing & 153 / 2 \\
Change in the annual mean of the departmental working-age population due to smoothing, maximum & 0.017\% \\
Change in a monthly cell of the departmental working-age population due to smoothing: median / 99th percentile / maximum & 0.18\% / 2\% / 12.7\% \\
\bottomrule
\end{tabularx}
\end{table}

The reconstructed history has short-run variability comparable to that of the observed period, which is the intended effect of the generated dynamics. The ICML total is slightly less volatile than the national reference because the noise of the 33 territories is independent and partly cancels in the sum. The largest change produced by the smoothing of territorial denominators (12.7\%) occurs in six cells of Choc\'o in 2020, where the unsmoothed series had a spike.

\subsection{National series}

\begin{figure}[H]
\centering
\includegraphics[width=0.92\textwidth]{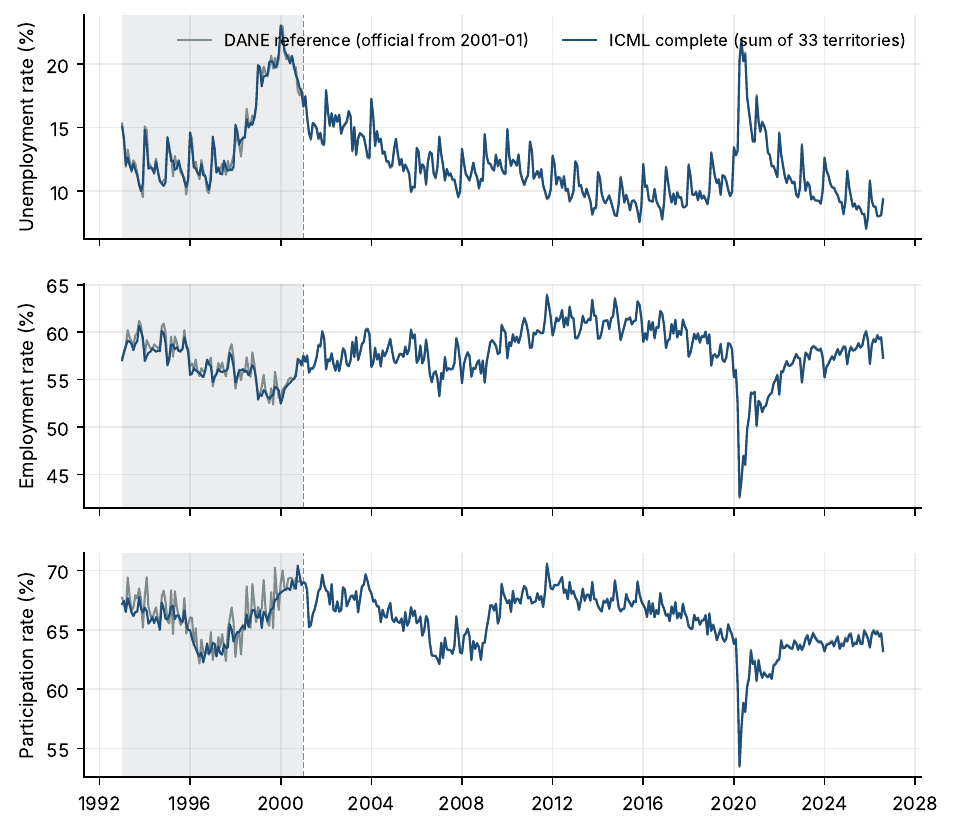}
\caption{National unemployment, employment and participation rates, January 1993 to August 2026. The shaded area is the reconstructed period (1993--2000), for which the reference series is not an official observation.}
\label{fig:national-rates}
\end{figure}

\begin{figure}[H]
\centering
\includegraphics[width=0.92\textwidth]{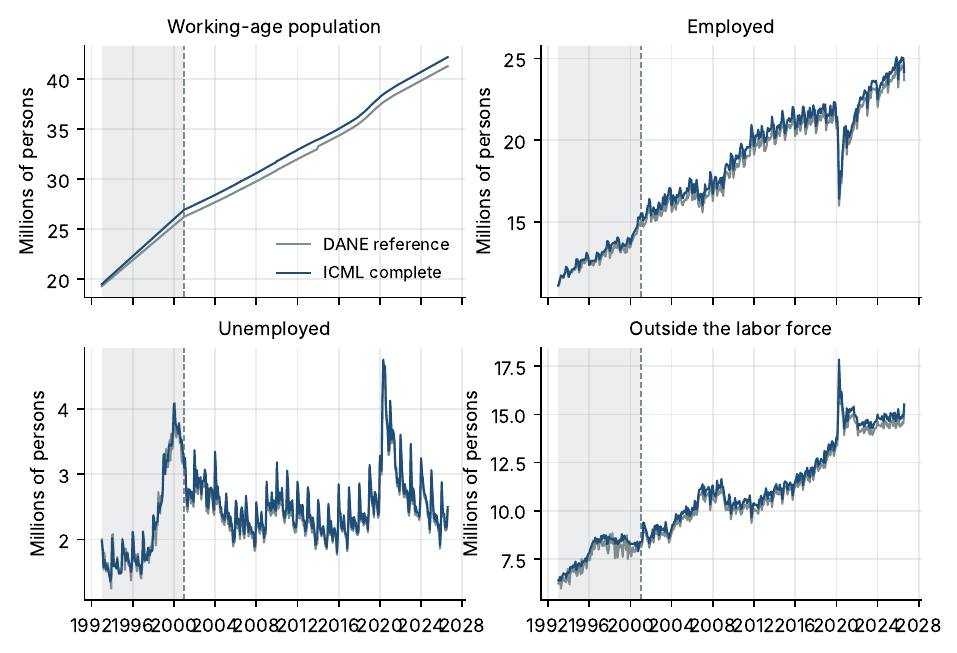}
\caption{National levels of the working-age population, employed, unemployed and population outside the labor force. Shaded area: reconstructed period.}
\label{fig:national-levels}
\end{figure}

Figure~\ref{fig:national-rates} shows the three rates for the two national series and Figure~\ref{fig:national-levels} the main levels. From January 2001 the DANE reference is the published series; the ICML total differs from it by the population the survey does not represent, and the rates of both are nearly identical because the unrepresented population is small and is assigned ratios close to the national ones. Table~\ref{tab:national} reports annual values.

\begin{table}[H]
\centering
\caption{National annual values. Levels are monthly means in millions of persons for the ICML total; rates are ratios of annual sums, in per cent. Informality rates for the census and GEIH universes.}
\label{tab:national}
\scriptsize
\setlength{\tabcolsep}{3.5pt}
\begin{tabular}{lrrrrrrrrrrr}
\toprule
 & \multicolumn{6}{c}{\textbf{ICML total (census universe)}} & \multicolumn{3}{c}{\textbf{DANE reference}} & \multicolumn{2}{c}{\textbf{Informality}} \\
\cmidrule(lr){2-7}\cmidrule(lr){8-10}\cmidrule(lr){11-12}
\textbf{Year} & PET & Empl. & Unemp. & $\TGP$ & $\TO$ & $\TD$ & $\TGP$ & $\TO$ & $\TD$ & Census & GEIH \\
\midrule
1993 & 19.93 & 11.73 & 1.59 & 66.9 & 58.9 & 12.0 & 67.4 & 59.3 & 12.0 & 68.9 & 68.8 \\
1996 & 22.72 & 12.71 & 1.69 & 63.4 & 56.0 & 11.7 & 64.0 & 56.4 & 11.8 & 68.9 & 68.8 \\
2000 & 26.44 & 14.52 & 3.66 & 68.8 & 54.9 & 20.1 & 69.1 & 55.2 & 20.1 & 69.7 & 69.5 \\
2001 & 27.16 & 15.62 & 2.77 & 67.7 & 57.5 & 15.0 & 67.7 & 57.6 & 15.0 & 69.6 & 69.4 \\
2005 & 29.20 & 16.96 & 2.29 & 65.9 & 58.1 & 11.9 & 66.0 & 58.2 & 11.8 & 67.4 & 67.3 \\
2010 & 32.05 & 19.08 & 2.61 & 67.7 & 59.5 & 12.0 & 67.7 & 59.6 & 12.0 & 66.6 & 66.4 \\
2015 & 34.72 & 21.27 & 2.15 & 67.4 & 61.2 & 9.2 & 67.5 & 61.3 & 9.2 & 62.2 & 62.0 \\
2019 & 37.69 & 21.75 & 2.65 & 64.7 & 57.7 & 10.8 & 64.8 & 57.7 & 10.9 & 60.2 & 60.0 \\
2020 & 38.57 & 19.45 & 3.83 & 60.4 & 50.4 & 16.5 & 60.4 & 50.4 & 16.5 & 60.2 & 60.0 \\
2021 & 39.26 & 20.82 & 3.32 & 61.5 & 53.0 & 13.8 & 61.5 & 53.1 & 13.8 & 59.4 & 59.2 \\
2022 & 39.84 & 22.49 & 2.83 & 63.6 & 56.4 & 11.2 & 63.6 & 56.5 & 11.2 & 57.9 & 57.7 \\
2023 & 40.41 & 23.26 & 2.63 & 64.0 & 57.5 & 10.2 & 64.1 & 57.6 & 10.2 & 56.4 & 56.1 \\
2024 & 41.01 & 23.51 & 2.66 & 63.8 & 57.3 & 10.2 & 63.9 & 57.4 & 10.2 & 55.9 & 55.7 \\
2025 & 41.58 & 24.32 & 2.37 & 64.2 & 58.5 & 8.9 & 64.3 & 58.6 & 8.9 & 55.7 & 55.5 \\
\bottomrule
\end{tabular}
\end{table}

The series reproduce the known episodes of the period: the rise of unemployment to about 20\% in 2000, its gradual decline to about 9\% in 2015, the contraction of 2020, when the employment rate fell to 50.4\% and unemployment rose to 16.5\%, and the recovery to an unemployment rate of 8.9\% in 2025. For years before 2001 these values belong to the reconstruction and inherit its assumptions.

In the reference vintage, the factor-weighted sum of the departments differs from the published national total by at most 0.05\% in population and 0.19\% in employed from 2014, and by up to 0.9\% between 2001 and 2013, when the capitals of the nine territories carry a geographic share that the national total of the survey did not yet include. That sum is reported as a diagnostic; the national figure of the GEIH universe is the published one.

Both universes are delivered for the 33 territories and for every month of the period, and in the GEIH universe each department-month carries its coverage status. In the census universe all quantities and rates are populated throughout. In the GEIH universe, the rates of the nine partially represented territories are left empty between 2001 and 2013, because the survey did not represent them in those years; the empty cells are part of the design of the product and are documented with it, and from 2014 those territories enter with the factor of their capital.

\subsection{Departmental patterns}

\begin{figure}[H]
\centering
\includegraphics[width=0.9\textwidth]{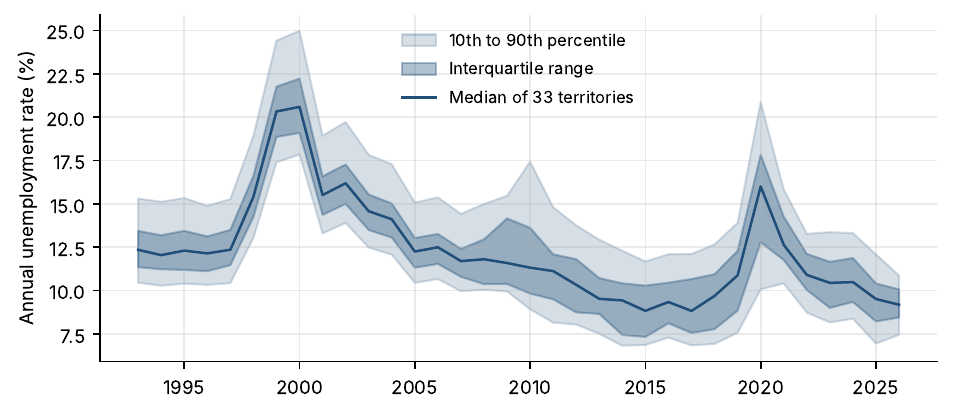}
\caption{Annual unemployment rate across the 33 territories: median, interquartile range and 10th--90th percentile band.}
\label{fig:dispersion}
\end{figure}

Figure~\ref{fig:dispersion} summarizes the distribution of annual unemployment rates across the 33 territories. The median territory follows the national cycle, with 20.6\% in 2000, 8.8\% in 2015, 16.0\% in 2020 and 9.5\% in 2025. Dispersion widens in downturns: the interquartile range is 3.1 points in 2000 and 5.1 in 2020, against 2.2 in 2025, and the range between the lowest and highest territory in 2025 (6.0\% to 12.8\%) is the narrowest of the selected years.

\begin{figure}[H]
\centering
\includegraphics[width=0.98\textwidth]{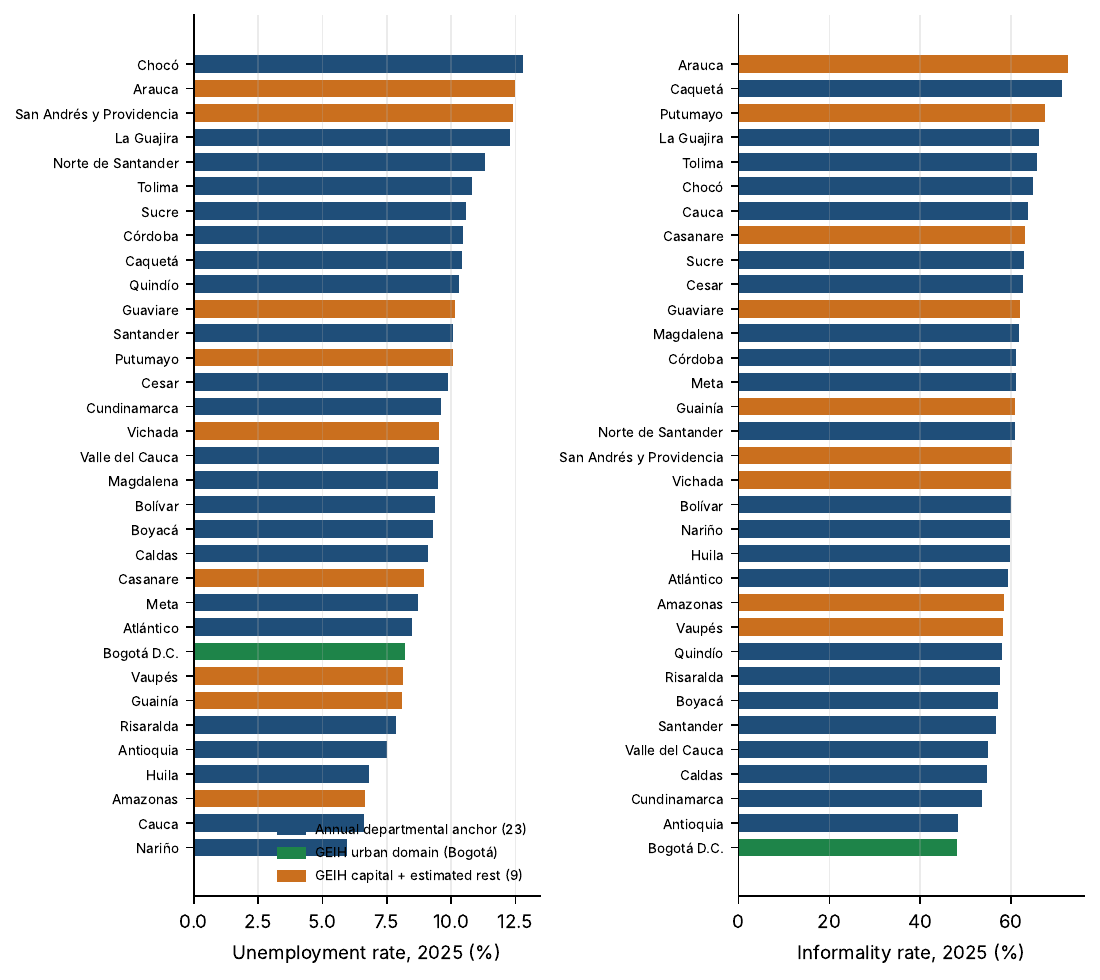}
\caption{Unemployment and informality rates by territory, 2025 (ratios of annual sums, census universe). Colors indicate the evidence behind the labor levels of each territory.}
\label{fig:departments}
\end{figure}

Table~\ref{tab:departments} and Figure~\ref{fig:departments} report the 2025 values by territory together with the type of evidence behind each one. Among the territories with an annual anchor, the unemployment rate ranges from 6.0\% (Nari\~no) to 12.8\% (Choc\'o) and the informality rate from 48.2\% (Antioquia) to 71.3\% (Caquet\'a). Bogot\'a combines the highest participation rate (70.5\%) with one of the lowest informality rates (48.1\%). For the nine territories built from a capital and an estimated rest, the values depend on the rule of Section~\ref{sec:representativeness} and should be read with the caveats of Section~\ref{sec:rest-evidence}.

\begin{table}[H]
\centering
\caption{Territorial indicators, 2025, census universe. Population in thousands (annual mean); rates in per cent (ratios of annual sums). Evidence: A = annual departmental anchor; U = GEIH urban domain; C = GEIH capital plus estimated rest of territory.}
\label{tab:departments}
\scriptsize
\begin{tabular}{llcrrrrr}
\toprule
\textbf{Code} & \textbf{Territory} & \textbf{Evid.} & \textbf{Population} & $\TGP$ & $\TO$ & $\TD$ & \textbf{Informality} \\
\midrule
05 & Antioquia & A & 7,199 & 64.3 & 59.4 & 7.5 & 48.2 \\
08 & Atlántico & A & 2,894 & 63.4 & 58.0 & 8.5 & 59.3 \\
11 & Bogotá D.C. & U & 8,069 & 70.5 & 64.7 & 8.2 & 48.1 \\
13 & Bolívar & A & 2,301 & 58.2 & 52.8 & 9.4 & 60.0 \\
15 & Boyacá & A & 1,284 & 56.5 & 51.2 & 9.3 & 57.1 \\
17 & Caldas & A & 1,066 & 59.3 & 53.9 & 9.1 & 54.6 \\
18 & Caquetá & A & 433 & 58.3 & 52.2 & 10.4 & 71.3 \\
19 & Cauca & A & 1,552 & 66.2 & 61.8 & 6.6 & 63.7 \\
20 & Cesar & A & 1,394 & 58.9 & 53.0 & 9.9 & 62.7 \\
23 & Córdoba & A & 1,892 & 64.7 & 58.0 & 10.4 & 61.2 \\
25 & Cundinamarca & A & 3,792 & 69.9 & 63.2 & 9.6 & 53.5 \\
27 & Chocó & A & 566 & 38.8 & 33.8 & 12.8 & 64.8 \\
41 & Huila & A & 1,166 & 56.8 & 52.9 & 6.8 & 59.8 \\
44 & La Guajira & A & 1,042 & 62.0 & 54.4 & 12.3 & 66.2 \\
47 & Magdalena & A & 1,502 & 60.2 & 54.5 & 9.5 & 61.7 \\
50 & Meta & A & 1,104 & 64.4 & 58.8 & 8.7 & 61.1 \\
52 & Nariño & A & 1,637 & 74.2 & 69.8 & 6.0 & 59.9 \\
54 & Norte de Santander & A & 1,674 & 59.6 & 52.9 & 11.3 & 60.8 \\
63 & Quindío & A & 593 & 60.0 & 53.9 & 10.3 & 58.1 \\
66 & Risaralda & A & 1,008 & 60.5 & 55.7 & 7.9 & 57.7 \\
68 & Santander & A & 2,372 & 62.8 & 56.5 & 10.1 & 56.7 \\
70 & Sucre & A & 1,000 & 63.6 & 56.9 & 10.6 & 62.8 \\
73 & Tolima & A & 1,357 & 62.0 & 55.3 & 10.8 & 65.8 \\
76 & Valle del Cauca & A & 4,692 & 63.4 & 57.3 & 9.5 & 55.0 \\
81 & Arauca & C & 321 & 62.4 & 54.6 & 12.5 & 72.6 \\
85 & Casanare & C & 482 & 64.3 & 58.5 & 8.9 & 63.2 \\
86 & Putumayo & C & 394 & 59.7 & 53.7 & 10.1 & 67.4 \\
88 & San Andrés y Providencia & C & 62 & 67.1 & 58.8 & 12.4 & 60.3 \\
91 & Amazonas & C & 87 & 59.7 & 55.8 & 6.6 & 58.5 \\
94 & Guainía & C & 59 & 57.3 & 52.7 & 8.1 & 60.8 \\
95 & Guaviare & C & 103 & 69.2 & 62.2 & 10.1 & 62.0 \\
97 & Vaupés & C & 49 & 59.9 & 55.0 & 8.1 & 58.2 \\
99 & Vichada & C & 127 & 61.6 & 55.7 & 9.5 & 60.1 \\
\bottomrule
\end{tabular}
\end{table}

\subsection{Informality}

\begin{figure}[H]
\centering
\includegraphics[width=0.9\textwidth]{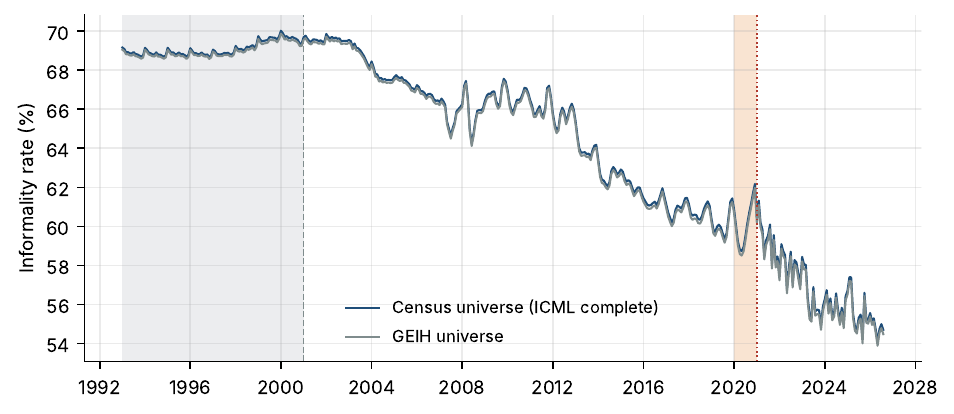}
\caption{National informality rate in the census and GEIH universes. Gray band: period without an official anchor. Orange band: 2020, filled with the interpolated scenario. Dotted line: change of definition in 2021.}
\label{fig:informality}
\end{figure}

Figure~\ref{fig:informality} shows the national informality rate in both universes. The rate is close to 69\% throughout the reconstructed period, declines from 69.6\% in 2001 to 60.2\% in 2019 and reaches 55.7\% in 2025 in the census universe (55.5\% in the GEIH universe). The level before 2021 depends on the bridge between definitions and the level before 2001 on a backcast; the uncertainty class attached to each record reflects this.

On held-out cities, the residual network has a mean absolute error of 5.5 percentage points under the earlier definition (2007--2019) and 11.4 points under the current one (2023--2025). A ridge regression obtains 9.5 and 10.7, and the mean of the training cities 8.7 and 9.6. The network is clearly better than the simple references in the earlier regime and not better in the current one, where the labeled sample is short. These errors are measured on urban labels, which are themselves latent reconstructions; they are not errors with respect to departmental truth.

The calibration \eqref{eq:kl-solution} is large. In the reference vintage, the employment-weighted mean absolute difference between the calibrated rate and the prior is 16.0 percentage points on average over the 404 months (between 8.6 and 21.4 points depending on the month), and the shift is upward in every period: 15.5 points in 1993--2000, 19.7 in 2001--2006, 16.0 in 2007--2019 and 13.1 in 2021--2026. The priors are learned from cities, where informality is lower than in the country as a whole, and the national target includes the rural population. The departmental estimates should therefore be read as follows: the \emph{ordering and relative distances} between departments come from an urban-trained model applied to departmental labor features, and the \emph{level} comes from the national target. The identities are exact (maximum residual of $10^{-5}$ persons) and the closure reproduces the national count within $10^{-8}$ persons.

The national rate derived for the GEIH universe is below the calibrated census-universe target by 0.16 points on average (between 0.09 and 0.25).

The estimated break at January 2001 is 2.47 percentage points for the national rate (standard error 0.08; 0.13 after adjusting for the autocorrelation of the residuals, which is 0.53). The estimate is stable for windows of 18 to 36 months (2.51, 2.47 and 2.56 points) and higher with 12 months (2.77), because the last months of the backcast contain a decline that is not information. After the splice, the one-month step of the national rate is 0.32 points instead of 2.78.

\subsection{Employment by sector}

\begin{table}[H]
\centering
\caption{Evaluation of the territorial sectoral predictor: mean and maximum absolute error per department-sector cell, in percentage points of the departmental employment share. LODO: leave-one-department-out.}
\label{tab:sector-metrics}
\footnotesize
\begin{tabular}{lrrrr}
\toprule
 & \multicolumn{2}{c}{\textbf{LODO, 2015--2024}} & \multicolumn{2}{c}{\textbf{2025 held out}} \\
\cmidrule(lr){2-3}\cmidrule(lr){4-5}
\textbf{Model} & Mean & Maximum & Mean & Maximum \\
\midrule
Ensemble (ridge and boosted trees, equal weights) & 1.171 & 21.04 & 1.277 & 17.44 \\
Ridge baseline & 1.236 & 25.12 & 1.370 & 23.60 \\
Boosted trees, 14 independent models & 1.284 & 21.58 & 1.330 & 14.18 \\
\bottomrule
\end{tabular}
\end{table}

The ensemble improves on the ridge baseline by about 0.07 percentage points per cell, a modest gain, and individual cells can be wrong by many points. After reconciliation, the product differs from the annual departmental shares of the survey by a mean of 0.20 points (maximum 3.1). For Bogot\'a, the external error before anchoring is 1.87 points; the 0.09 points obtained after anchoring measure fit and not prediction. The largest monthly adjustment that the reconciliation imposes on a prior is 5.7 points, in 2020. The sensitivity of the backcast to its assumption is large: carrying the 2015 territorial structure back to 1993 without attenuation, instead of with a half-life of twelve years, changes one cell by up to 23 points. Before 2015 the departmental sector structure should be regarded as an assumption with a declared sensitivity.

\begin{figure}[H]
\centering
\includegraphics[width=0.92\textwidth]{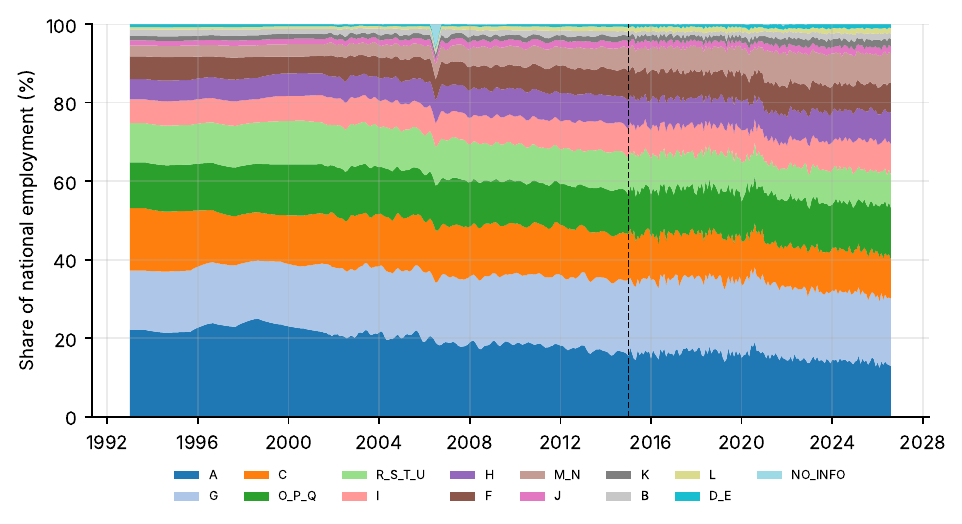}
\caption{National composition of employment by sector, January 1993 to August 2026 (census universe). The dashed line marks the start of calendar-month observations (2015).}
\label{fig:sectors}
\end{figure}

At the national level (Figure~\ref{fig:sectors} and Appendix~\ref{app:sectors}), the composition shifts from agriculture and manufacturing, which fall from 22.1\% and 15.9\% of employment in 1993 to 14.2\% and 11.0\% in 2025, toward trade (15.1\% to 17.2\%), transportation (5.2\% to 7.6\%) and professional and administrative services (2.8\% to 7.6\%). Shares before 2015 depend on the bridges between classifications.

\subsection{Internal consistency of the delivered products}

In the reference vintage, the labor identities hold with a maximum residual below $10^{-8}$ persons; the sum of the 33 departments equals the ICML national total within $10^{-8}$ persons; the employment of the informality product equals that of the labor product within $10^{-5}$ persons; and employment by sector adds to departmental employment within $10^{-4}$ persons and to the national sector totals within $10^{-8}$ persons. The delivery passes the 52 checks.

\section{Limitations}
\label{sec:limitations}

The limitations are stated by product and by period, since they differ substantially.

\begin{itemize}
  \item Status of the results: the series are research estimates and not official statistics. The system records, in every run, evidence checks that cannot pass with the available data, and it does not authorize publication on its own. An independent scientific validation of the informality closure has not been carried out.
  \item History before 2001: the months from 1993 to 2000 are reconstructed from annual survey levels of a different design. Their short-run dynamics (noise and seasonality) are generated to match the statistical behavior of the later period and carry no month-specific information. The annual means of the earlier survey are no longer reproduced for the working-age population, by design. The seam at January 2001 leaves an employment change of 2.8\% in one month in the ICML total, which is within the normal monthly variation of the series (median 1.4\%, 99th percentile 5.2\%) but is not removed. The universe of the national reference before 2001 is not verified against the territorial projections.
  \item Territories without an annual anchor: for the nine partially represented territories, the labor levels outside the capital are derived from ratios observed in the rest of territory of other departments, which is Andean and Caribbean. No survey observes the rural Amazon and Orinoco regions. The validation of Section~\ref{sec:rest-evidence} gives errors of 4.5--7.4 percentage points when the capital holds less than a fifth of the population, which is the situation of Putumayo and Vichada. The urban signals of the three smallest capitals depend on regularization choices: their latent months are poorly conditioned, and their windows are reproduced within a tolerance rather than exactly.
  \item Urban histories before 2021: nine capitals have no published urban history before 2021 and their 2007--2020 trajectories are those of donor cities. The donor rule was validated on cities with known history, but the trajectories remain borrowed.
  \item Latent months: monthly values derived from moving windows are not observations. They are exact with respect to the published windows and otherwise determined by a prior and a smoothness criterion. Months inside the open window are provisional and are revised with each publication.
  \item Supervised models: the transfer evaluation is retrospective: predictors and labels use the latest vintage, hyperparameters were chosen with all departments, and anchored departments differ systematically from unanchored ones. After reconciliation the departmental model does not improve on the urban signal. Estimates that depend on the boosted model change across minor releases of the library and across platforms by amounts that are not negligible for small territories; the reference is reproducible only under the pinned release, and a predictor that is stable across versions has not been evaluated.
  \item Informality: departmental informality is never observed. The priors transfer an urban model to departments and the national calibration shifts them by 13--20 percentage points on average, so the departmental level is essentially determined by the national target and the departmental pattern by the urban model. The rate of the unrepresented part of a territory is assumed equal to that of its represented part. The level before 2021 depends on a bridge between two operational definitions, the months of 2020 are a scenario, and the period before 2007 rests on six aggregate anchors and a backcast. The configuration of the deep-learning library used to train the priors is not fully recorded, which affects the ability to retrain them identically, not their reuse. No numerical criterion yet decides when frozen priors have drifted too far from new data.
  \item Sectoral employment: the product is experimental. Departmental sector labels are annual and exist for 23 departments from 2015; nine territories have no observed sectoral anchor of any kind. The territorial structure before 2015 is an assumption whose sensitivity reaches 23 points in a cell. The split between mining and utilities relies on a fixed ratio. The annual anchors were taken from an earlier extraction and have not been collated again with the annexes.
  \item Anchors are bounded, not exact: the adopted allocation of the national residual moves annual departmental figures by up to 1.9\%. The amount absorbed by each cell is published, but users who require exact official annual figures should take them from the source.
  \item Operation after the frozen cutoff: the rules that extend the frozen parameters beyond August 2026 are implemented and unit-tested, but a complete run with real annexes of a later month had not been executed when this paper was written.
  \item Uncertainty: the products do not include interval estimates. The evidence classes, the deviation from official figures, the transfer errors and the size of each adjustment are reported instead, and they are not a substitute for a probabilistic statement.
\end{itemize}

\section{Conclusions}
\label{sec:conclusions}

This paper has described a framework that reconstructs monthly labor-market, informality and sectoral-employment series for the 33 first-level territories of Colombia from 1993 to 2026, and the system that produces, audits and updates them.

Three results are worth stressing. First, the gap between departmental aggregates and the national total of the household survey is a matter of coverage and not of estimation: with official figures alone, the national total equals 23 departments, Bogot\'a and nine capitals within two hundredths of a per cent in population. Publishing every territory for the whole territory and for the represented part, linked by explicit factors, removes the need to distort either the departments or the national total. Second, the official domains are not additive, and the choice of who absorbs the difference matters more than the choice of model: keeping annual anchors exact distorts other official figures by 9--12\%, whereas a proportional allocation keeps every official figure within 1.9\% and makes the absorbed amount visible. Third, a supervised model that looks best before reconciliation is not the best after it, and after reconciliation it does not improve on the survey's own urban domains. Evaluating models in the form in which their output is published changed the design of the product.

For the territories with an official reference, the published figures deviate from it by a median of 0.04--0.43\% and by at most 1.89\%. For territories treated as unobserved, the error on labor rates is about 2.2 percentage points. For the territories that are in fact unobserved, and for departmental informality and sectoral composition in general, no direct measure of accuracy exists; the paper reports instead what each estimate rests on and how much each adjustment moves it.

The operational contribution is a set of rules that make the panel maintainable: a month is processed only with both official publications; closed years do not change in an ordinary update; frozen parameters are reused and their extension is tagged; a reference vintage must be reproduced by any recomputation; and every delivery is verified against identities that hold across products and universes.

Several extensions follow from the limitations. The most useful would be an independent validation of the informality closure and a criterion for the drift of its frozen priors; interval estimates, at least for the anchored territories; a predictor for provisional months that is stable across library versions; and the verification of the universe of the national reference before 2001. More fundamentally, the framework shows the limit of what published aggregates allow: model-based reconstruction produces coherent series for territories that the survey does not cover, but it does not replace observation there.

\section*{Ethical Statement}

This study uses only aggregated, anonymized and publicly available data from official statistical sources. No individual-level, confidential or personally identifiable records were accessed or processed, and no ethical approval was required.

\section*{Acknowledgments}

The author acknowledges the work of the Departamento Administrativo Nacional de Estad\'istica (DANE) in producing and publishing the statistics on which this study relies. The views and conclusions expressed here are solely those of the author. The results are research estimates and are not official statistics.

\section*{Funding and Competing Interests}

This research received no funding from public agencies, private institutions or nonprofit organizations. The author declares no competing interests.

\section*{Data and Code Availability}

The source code, the input annexes, the frozen parameters, the reference vintage, the tests and the documentation are available under the MIT license (version 2.6.1) at\\ \url{https://github.com/jaimevera1107/colombian-labor-market}. Deliveries are generated locally from the repository with the commands of Section~\ref{sec:system}; the integrity of a copy can be verified against the hash listing distributed with it.
The reconstructed series of release 2.6.1, with data through August 2026, are archived at \url{https://doi.org/10.5281/zenodo.24357243}.

\bibliographystyle{plainnat}
\bibliography{references}

\clearpage
\appendix
\section{Parameters of the Reference Vintage}
\label{app:parameters}

Table~\ref{tab:parameters} collects the parameters quoted in the text. All of them are read from the versioned configuration files of the reference vintage; none is tuned at run time.

\begin{longtable}{L{3.4cm} L{7.2cm} R{3.6cm}}
\caption{Configured parameters by stage.}\label{tab:parameters}\\
\toprule
\textbf{Stage} & \textbf{Parameter} & \textbf{Value} \\
\midrule
\endfirsthead
\toprule
\textbf{Stage} & \textbf{Parameter} & \textbf{Value} \\
\midrule
\endhead
\bottomrule
\endfoot
National & First observed month & 2001-01 \\
National & Annual splice, overlapping periods & 2 \\
National & Disaggregation method & Chow--Lin, maximum likelihood \\
National & Relative uncertainty of historical anchors & 0.05 \\
National & Minimum anchor uncertainty & 10,000 persons \\
National & Maximum annual / monthly relative shift & 0.10 / 0.15 \\
National & Seasonal template window & 2004--2019 \\
National & Seasonal transfer strength & 0.5 \\
National & STL seasonal / trend windows & 13 / 19 \\
\midrule
Urban & Donor selection years / holdout year & 2021--2024 / 2025 \\
Urban & History assigned from donors & 2007--2020 \\
Urban & Tie-break weights (five criteria) & 0.30, 0.20, 0.20, 0.10, 0.20 \\
Urban & Closure ramp & 2015--2020 \\
Urban & Splice overlap & 12 months \\
\midrule
Departmental & Anchored years & 2007--2025 \\
Departmental & Anchored departments & 23 \\
Departmental & Maximum closure change of an anchor & 2.01 persons \\
Departmental & Anchor tolerance of the reconstruction & $10^{-4}$ persons \\
Departmental & Population backcast & 1993--2006 \\
\midrule
Integration & Number of 12-month windows & 57 \\
Integration & Latent months & 2021-01 to 2026-08 \\
Integration & Smoothness / bridge weights & 0.65 / 0.12 \\
Integration & Relative floor & 0.10 \\
Integration & Maximum monthly ratio (unemployment) & 2.0 \\
Integration & Maximum window tolerance, flexible cities & 0.021 \\
\midrule
Modeling & Training years & 2007--2025 \\
Modeling & Lags & 1 and 12 months \\
Modeling & Folds & 5 groups $\times$ 3 windows \\
Modeling & Embargo & 1 year \\
Modeling & Ridge penalty & 25 \\
Modeling & Trees / depth / learning rate & 200 / 3 / 0.045 \\
Modeling & Minimum child weight & 12 \\
Modeling & Row / column subsampling & 0.9 / 0.9 \\
Modeling & $L_2$ / $L_1$ penalties & 8.0 / 0.05 \\
Modeling & Maximum number of clusters per fold & 5 \\
Modeling & Bridge taper for provisional months & 12 months \\
\midrule
Reconciliation & Reconciled to the national total from & 2001-01 \\
Reconciliation & Capitals inside the national total from & 2014-01 \\
Reconciliation & Residual allocation & proportional \\
Reconciliation & Closure tolerances & $10^{-3}$ persons \\
\midrule
Reconstruction & Splice month & 2001-01 \\
Reconstruction & Noise scale & 0.45 \\
Reconstruction & STL period / seasonal window & 12 / 13 (robust) \\
Reconstruction & Seasonal reference period & 2001--2018 \\
\midrule
Quality & Accounting tolerance & $10^{-5}$ persons \\
Quality & Partial-factor knots & 2001-01, 2014-01, last anchored December \\
\midrule
Informality & Splice window & 24 months \\
Informality & Network width / residual blocks / dropout & 128 / 1 / 0.1 \\
Informality & Beta precision & 20 \\
Informality & Learning rate / weight decay & $10^{-3}$ / $10^{-3}$ \\
Informality & Epochs / batch size / patience & 160 / 512 / 18 \\
Informality & Calibration tolerance & $10^{-4}$ persons \\
\end{longtable}

\section{Products and Their Columns}
\label{app:products}

\begin{table}[H]
\centering
\caption{The eight internal products.}
\label{tab:products}
\footnotesize
\begin{tabularx}{\textwidth}{L{3.2cm} L{2.6cm} Y}
\toprule
\textbf{Product} & \textbf{Rows at 2026-08} & \textbf{Content} \\
\midrule
Labor, departments & 13,332 & Six levels and three rates of the whole territory; represented levels and factors per quantity; persons of the national residual absorbed per component; source of the levels; status of the factor \\
Labor, cities & 13,332 & Six levels of 33 non-additive urban signals with origin and source status \\
Labor, nationals & 808 & Two series per month: ICML total and DANE reference, with reference status and an indicator of official observation \\
Informality, departments & 13,332 & Prior and calibrated rate; informal and formal employment in both universes and outside the represented universe; provenance, uncertainty class and estimation status \\
Informality, cities & 13,332 & Latent rate and informal and formal employment of 33 urban signals \\
Informality, nationals & 808 & Informal and formal employment and rate in both universes; provenance and uncertainty class \\
Sectoral, departmental & 199,980 & Employment and share by department, sector and month in both universes; status of the prior and of the anchors \\
Sectoral, national & 6,060 & Employment and share by sector and month; source status \\
\bottomrule
\end{tabularx}
\end{table}

A delivery contains each product once per universe (16 files). Territory codes are two-digit text and must be read as text to preserve leading zeros.

\section{Sector Groups}
\label{app:sectors}

\begin{table}[H]
\centering
\caption{The 15 sector groups and their national employment shares (per cent, annual means, census universe).}
\label{tab:sectors}
\small
\begin{tabular}{llrr}
\toprule
\textbf{Code} & \textbf{Group} & \textbf{1993} & \textbf{2025} \\
\midrule
\code{A} & Agriculture, livestock, forestry and fishing & 22.1 & 14.2 \\
\code{B} & Mining and quarrying & 1.7 & 1.5 \\
\code{C} & Manufacturing & 15.9 & 10.9 \\
\code{D\_E} & Electricity, gas, water and waste management & 0.8 & 1.1 \\
\code{F} & Construction & 5.8 & 6.7 \\
\code{G} & Wholesale and retail trade; repair of vehicles & 15.1 & 17.2 \\
\code{H} & Transportation and storage & 5.2 & 7.6 \\
\code{I} & Accommodation and food services & 6.0 & 7.7 \\
\code{J} & Information and communications & 1.2 & 1.7 \\
\code{K} & Financial and insurance activities & 1.1 & 1.8 \\
\code{L} & Real estate activities & 0.6 & 1.4 \\
\code{M\_N} & Professional, scientific, technical and administrative services & 2.8 & 7.6 \\
\code{O\_P\_Q} & Public administration, education and health & 11.7 & 12.1 \\
\code{R\_S\_T\_U} & Arts, entertainment, other services and households & 10.0 & 8.4 \\
\code{NO\_INFO} & Not reported & 0.0 & 0.0 \\
\bottomrule
\end{tabular}
\end{table}

\section{Closure Variants Evaluated}
\label{app:closures}

\begin{table}[H]
\centering
\caption{Structural properties of the closure variants on the production panel (nothing hidden).}
\label{tab:closures}
\footnotesize
\begin{tabularx}{\textwidth}{L{3.6cm} L{3.4cm} Y L{2.6cm}}
\toprule
\textbf{Variant} & \textbf{Maximum anchor deviation} & \textbf{Sum of 33 territories against the national survey total} & \textbf{Bogot\'a labor population against its projection} \\
\midrule
Anchors exact, national survey total & 0 & exact & 18.1\% \\
Uniform raking & employed 3.1\%, population 7.1\% & exact & 4.8\% \\
Demographic universe & 0 & $+1.1$ to $+5.0\%$ (coverage gap) & 1.7\% \\
Components only & 0 & exact for components; population $+1.8$ to $+4.5\%$ & 0\% \\
\bottomrule
\end{tabularx}
\end{table}

\end{document}